# X-ray properties of X-CLASS-redMaPPer galaxy cluster sample: The luminosity-temperature relation

Mona Molham[1,2]⋆, Nicolas Clerc[3], Ali Takey[1,4], Tatyana Sadibekova[5,6], A. B. Morcos[1], Shahinaz Yousef[2], Z. M. Hayman[2], Maggie Lieu[7], Somak Raychaudhury[8], and Evelina R. Gaynullina[6]

[1] *National Research Institute of Astronomy and Geophysics (NRIAG), 11421 Helwan, Egypt*
[2] *Astronomy and Meteorology Department, Faculty of Science, Cairo University, 12613 Giza, Egypt*
[3] *IRAP, Université de Toulouse, CNRS, CNES, UPS, (Toulouse), France*
[4] *Hiroshima Astrophysical Science Center, Hiroshima University, 1-3-1 Kagamiyama, Higashi-Hiroshima, Hiroshima 739-8526, Japan*
[5] *Laboratoire AIM, CEA/DSM/IRFU/SAp, CEA Saclay, 91191 Gif-sur-Yvette, France*
[6] *Ulugh Beg Astronomical Institute of Uzbekistan Academy of Science, 33 Astronomicheskaya str., Tashkent, UZ-100052, Uzbekistan*
[7] *European Space Astronomy Centre, ESA, Villanueva de la Caada, E28691 Madrid, Spain*
[8] *Inter-University Centre for Astronomy and Astrophysics, Pune 411007, India*

**ABSTRACT**

This paper presents results of a spectroscopic analysis of the X-CLASS-redMaPPer (XC1-RM) galaxy cluster sample. X-CLASS is a serendipitous search for clusters in the X-ray wavebands based on the XMM-Newton archive, whereas redMaPPer is an optical cluster catalogue derived from the Sloan Digital Sky Survey (SDSS). The present sample comprises 92 X-ray extended sources identified in optical images within 1′ separation. The area covered by the cluster sample is ∼ 27 deg$^2$. The clusters span a wide redshift range ($0.05 < z < 0.6$) and 88 clusters benefit from spectrosopically confirmed redshifts using data from SDSS Data Release 14. We present an automated pipeline to derive the X-ray properties of the clusters in three distinct apertures: $R_{500}$ (at fixed mass overdensity), $R_{fit}$ (at fixed signal-to-noise ratio), $R_{300kpc}$ (fixed physical radius). The sample extends over wide temperature and luminosity ranges: from 1 to 10 keV and from $6{\times}10^{42}$ to $11{\times}10^{44}$ erg s$^{-1}$, respectively. We investigate the luminosity-temperature (L-T) relation of the XC1-RM sample and find a slope equals to $3.03 \pm 0.26$. It is steeper than predicted by self-similar assumptions, in agreement with independent studies. A simplified approach is developed to estimate the amount and impact of selection biases which might be affecting our recovered L-T parameters. The result of this simulation process suggests that the measured L-T relation is biased to a steeper slope and higher normalization.

**Key words:** X-rays: galaxies: clusters – galaxies: clusters: intracluster medium – galaxies: clusters: general

## 1 INTRODUCTION

Studies of galaxy clusters at high energies are developing rapidly over the last decade by harvesting the X-ray mission archives and thanks to dedicated X-ray missions targeting extragalactic sources over large portions of the sky. Clusters are the most massive luminous (∼$10^{43}$ to $10^{45}$ erg s$^{-1}$) gravitationally bound structures in the Universe. They are dominated by hot gas (intracluster medium), galaxies, active galactic nuclei (AGN), and dark matter. The hot intracluster medium (ICM) emits X-ray photons via free-free and line emission which makes them unambiguously detected as extended X-ray sources up to high redshifts. Galaxy clusters are established sensitive probes of the underlying cosmolog-ical model of our Universe (Reiprich & Böhringer 2002; Voit et al. 2005; Vikhlinin et al. 2009b; Rozo et al. 2010; Allen et al. 2011; Sehgal et al. 2011). One of the first X-ray catalogs of galaxy clusters was the EMSS catalog (Gioia et al. 1990), assembled by serendipitous searches in data acquired by the Einstein Observatory. Subsequently, many cluster samples were detected from the ROSAT mission, both in pointed observations and in its all-sky survey (Vikhlinin et al. 1998; Böhringer et al. 2000; Böhringer et al. 2004; Horner et al. 2008).

⋆ E-mail: monamolham@nriag.sci.eg

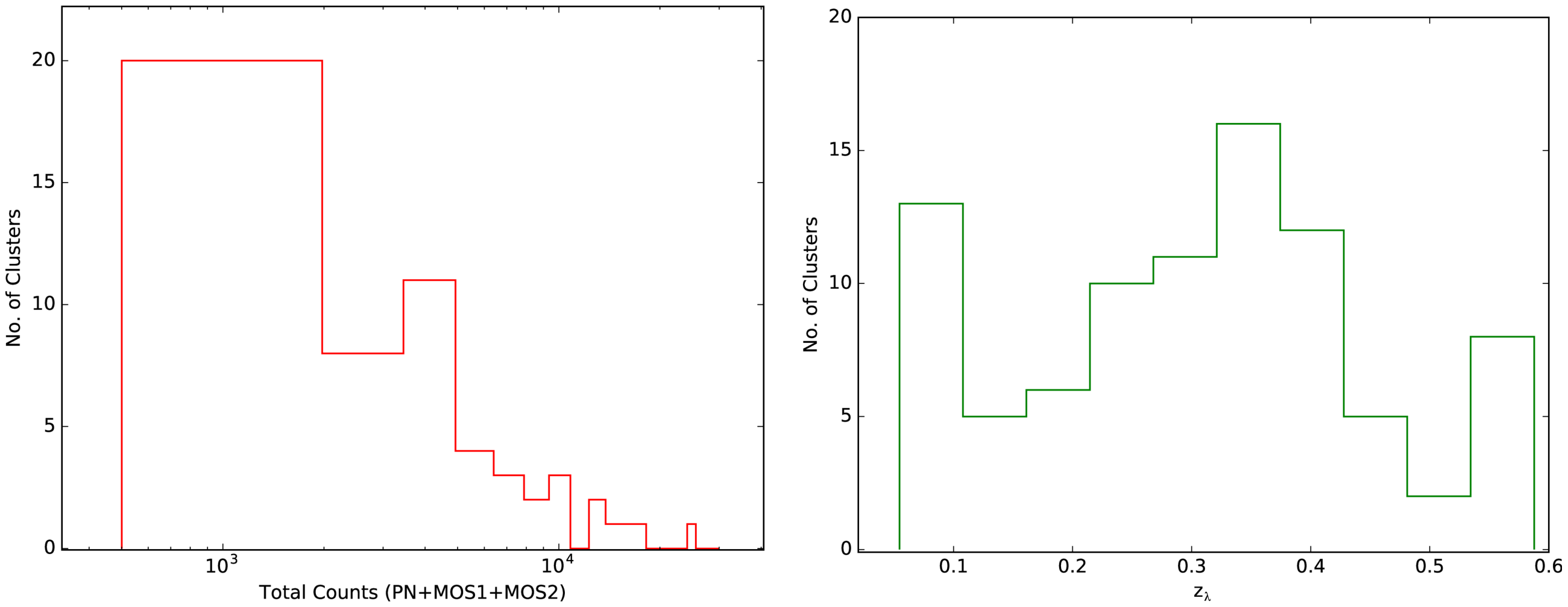

**Figure 1.** The distribution of the net counts and redshifts of the XC1-RM galaxy cluster sample. Left: The total counts (PN+MOS1+MOS2) distribution of the galaxy clusters in XC1-RM sample. Right: The photometric redshifts ($z_\lambda$) distribution of the galaxy clusters in XC1-RM sample.

With the advent of new-generation X-ray satellite missions like XMM-Newton and Chandra, an increased number of clusters were discovered in novel areas of the mass-redshift plane. New projects were triggered, among them numerous surveys exploiting their archive databases: for instance X-CLASS (Clerc et al. 2012), 2XMMi/SDSS (Takey et al. 2014; Takey et al. 2016), the XCS survey (Lloyd-Davies et al. 2011; Mehrtens et al. 2012), and dedicated connected wide-area surveys such as COSMOS (Finoguenov et al. 2007), XMM-BCS (Šuhada et al. 2012), the XMM-LSS (Pierre et al. 2004; Pacaud et al. 2006; Pacaud et al. 2007) and the XXL survey (Pierre et al. 2016; Pacaud et al. 2016).

Under the assumption that galaxy clusters are self-similar objects whose formation process is dominated by gravity, Kaiser (1986) found the correlation among X-ray cluster observable properties are described by a power law. The X-ray luminosity-temperature (L-T) relation is one of the most investigated such scaling relations (Pratt et al. 2009; Mittal et al. 2011; Maughan et al. 2012; Takey et al. 2019) and it is found that the slope of the relation is steeper than the self-similar prediction (which is equal to two). This is explained by non-gravitational physical processes, such as AGN heating and supernovae feedback. Determining the contribution of each process to the observed deviation from self-similarity calls for refined modelling and most often for numerical simulations. Furthermore, the inevitable Malmquist and Eddington biases are affecting the measurement of scaling relations, so that selection effects should be taken into consideration to fully understand their form and evolution (Pacaud et al. 2007; Mantz et al. 2010a; Mantz et al. 2010b). Malmquist bias in particular arises due to the fact that at greater distances, one can detect high luminosity sources in larger proportions than low luminosty ones. Therefore if a sample is limited in flux, higher luminosity sources appear over-represented. Several authors designed approaches correcting for the effects arising from different selection biases specific to a survey, e.g. Pratt et al. 2009; Vikhlinin et al. 2009a; Mantz et al. 2010b; Clerc et al. 2014; Lovisari et al. 2015; Bharadwaj et al. 2015; Zou et al. 2016; Giles et al. 2016.

In this paper, we present for the first time X-ray spectral properties of the XC1-RM galaxy cluster sample (Sadibekova et al. 2014) measured within three different apertures and the automated procedure that we embraced in this study. We investigate the observed L-T relation of the cluster sample and describe the simulation procedure adopted to probe the Malmquist bias effect on the observed L-T relation, then we interpret the results.

This paper is organized as follows. In section 2, we discuss the cluster sample, data reduction, data analysis, and derivation of the temperature of intra-cluster gas in different apertures. In section 3, we present the results of the L-T relation in the (0.5 – 2.0) keV energy band, bolometric luminosities, and the simulation approach for assessing the Malmquist bias affecting our observed L-T relation. Discussion and summary of the key results are provided in section 4 and 5, respectively. Throughout this paper, we assume a flat cosmological model with: the matter density $\Omega_m = 0.3$, the dark energy density $\Omega_\Lambda = 0.7$, and the Hubble constant $H_0 = 70$ km s$^{-1}$ Mpc$^{-1}$, unless stated otherwise.

## 2 GALAXY CLUSTER SAMPLE AND DATA PROCESSING

In Sect. 2.1, we describe the construction of the data sample. The procedure to determine the spectroscopic redshifts is described in Sect. 2.2, and we summarize the data reduction procedure in Sect. 2.3. The method to define clusters' emis-sion radii, and the spectral analysis process are presented in Sect. 2.4, and Sect. 2.5, respectively.

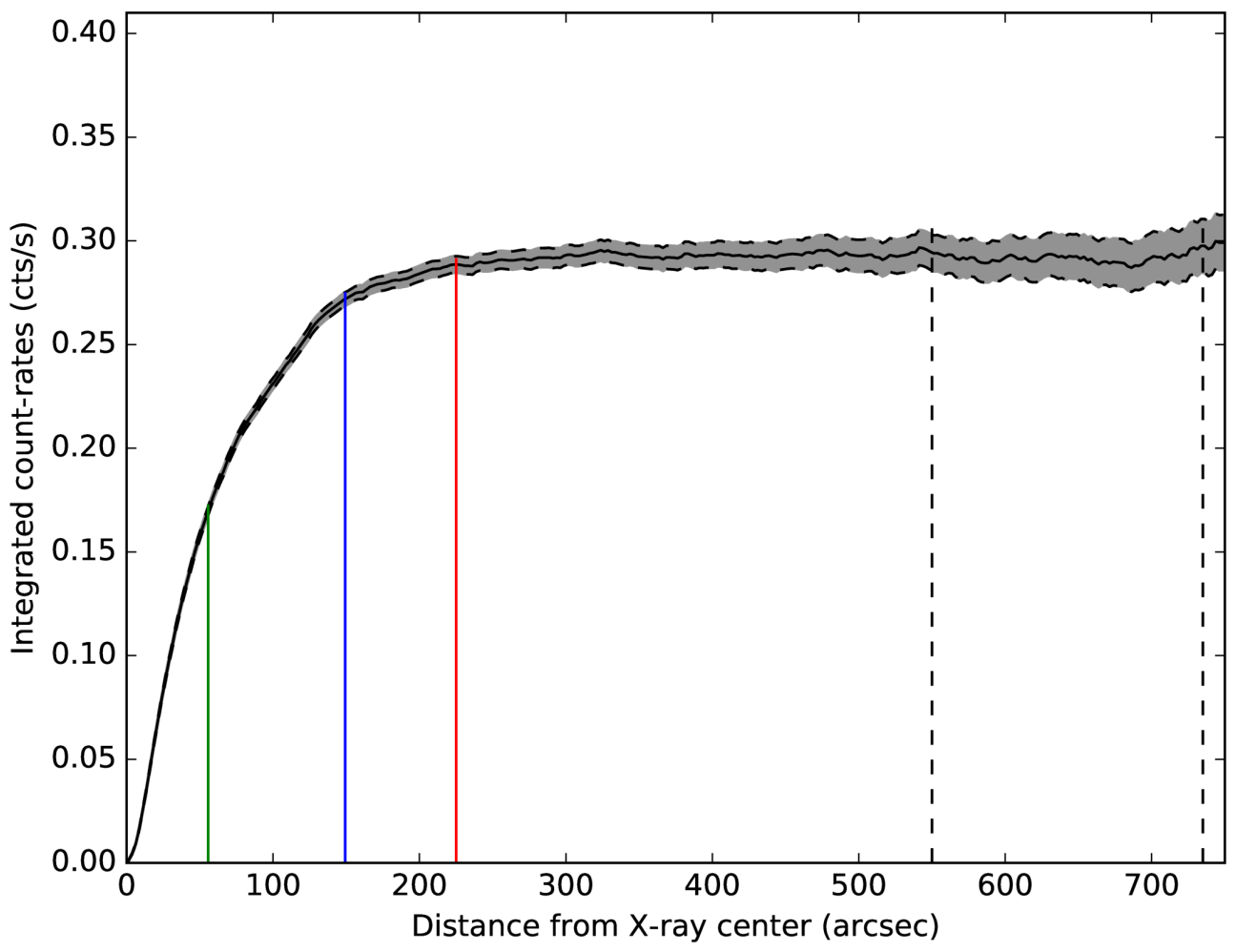

**Figure 2.** The net count rate (PN+MOS1+MOS2) growth curve of the cluster X-CLASS 377 and the gray area represents the 1 $\sigma$ error bar. $R_{300\mathrm{kpc}}$, $R_{500}$, and $R_{\mathrm{fit}}$ are represented by the green, blue, and red lines, respectively. The dashed vertical lines indicate the region we used for the background annulus.

### 2.1 Galaxy cluster sample

The X-CLASS-redMaPPer (XC1-RM) cluster sample (Sadibekova et al. 2014) is a joint sample between X-CLASS and redMaPPer catalogues in X-ray and optical bands, respectively using the best overlap between two survey footprints, which covers an area of 27 deg$^2$. The X-CLASS catalogue[§] (Clerc et al. 2012) embraces 845 class 1 (C1) X-ray selected galaxy clusters (C1 definition, Sect.2.3) detected in 2774 XMM archival observations by May 2010. The redMaPPer catalogue in the optical waveband is based on the red-sequence cluster finder algorithm which was applied to SDSS DR8 (Rykoff et al. 2014). For this study, we use the best joint X-optical sample of 92 clusters with the positional matching within an aperture of 1′. The accurate individual X-ray measurements have been provided for each cluster to assure the best source position and an optimal masking of other field detections. Then for the best flux estimate, a circular aperture around the X-ray source position was tuned in a semi-automatic manner from the net count rate (Clerc et al. 2012), where count rate (cts/s) is the mean number of photons collected by the telescope from the X-ray source in the direction of the optical axis in 1 second.

The cluster redshift $z_\lambda$ is photometric and provided from the redMaPPer catalogue. The ranges of redshifts and the total count rates spanned by our cluster sample are $0.05 < z < 0.6$ and 0.03 to 1.4 count rates, respectively. Fig. 1 shows the total counts and the photometric redshifts ($z_\lambda$) distribution of the clusters in our sample.

§ http://xmm-lss.in2p3.fr:8080/l4sdb/

### 2.2 Galaxy cluster sample with spectroscopic redshift in SDSS-DR14

The Sloan Digital Sky Survey fourteenth data (SDSS-DR14) was released on 31st July 2017. The important class of data for our study is the spectroscopic data, more than four million spectra, which are optical spectra from (SDSS/SEGUE/BOSS/SEQUELS/eBOSS). It is the first release of data from the SDSS component extended Baryon Oscillation Spectroscopic Survey (eBOSS), including spectra from the eBOSS subprograms (SPIDERS) survey (Abolfathi et al. 2018). We ran a query to search the SDSS-DR14 for spectroscopic redshifts ($z_s$) for our list of clusters. We searched for galaxies that have $z_s$ around each cluster X-ray position in a circle of radius equal 17′ which equals a physical radius of 1 Mpc at redshift 0.05. That step was made through SQL-based interface to the Catalog Archive Server (CAS) database, where we got a table contains RA, Dec, source name, $z_s$ and its uncertainty for all the galaxies. The first criterion aims to present all the possible results to a lower redshift limit of 0.05 for a physical radius of 1 Mpc, then we will identify the real members in each cluster. We found galaxies with $z_s$ around 90 clusters. Secondly, around each cluster X-ray position we accepted only galaxies with $z_s$ that are within a physical radius of 1 Mpc based on its photometric redshift ($z_\lambda$), and that justifies the condition of having $z_\lambda - 0.04(1 + z_\lambda) < z_s < z_\lambda + 0.04(1 + z_\lambda)$. This redshift range was suggested in (Wen et al. (2009),Takey et al. (2013)), and we used it to include most of galaxies with $z_s \approx z_\lambda$. After the previous step, the list shortened to just 88 clusters which have galaxies with $z_s$, and for each cluster we calculated the weighted average of the $z_s$ of its galaxies, where the weighted average= $\sum w z_s / \sum w$, $w = 1/(\Delta z_s)^2$. About 40 percent of the clusters have more than 10 galaxies with $z_s$. The relative change between $z_\lambda$ and $z_s$ is on average equal 0.0025 which it shows that $z_\lambda$ is in good agreement with $z_s$ (Table A1). Finally, we have 88 clusters with spectroscopic redshifts and for the remaining four clusters we used the photometric redshift $z_\lambda$ obtained from Sadibekova et al. (2014) since we did not find any spectroscopic redshifts $z_s$ in the Nasa/IPAC Extragalactic Database (NED).

### 2.3 Filtering and processing of XMM observations

The filtered event lists and other complementary files were created based on the XMM-LSS pipeline developed by Pacaud et al. (2006). The main steps are summarized in (Clerc et al. 2012) and we will recall them into the next points. (i) The calibrated event lists of the three EPIC cameras (PN, MOS1, and MOS2) were generated using the XMM-Newton "Scientific Analysis System" (SAS) tasks **emproc** and **epproc**. Then they were filtered from high-background periods to produce images. (ii) The created images were co-added and filtered in wavelet space, then sources were detected by running SEXTRACTOR (Bertin & Arnouts 1996) on them. (iii) The XAMIN pipeline (a maximum likelihood profile fitting procedure) provided the detected sources with number of parameters describing their properties. The C1 sample contains clusters characterized with the parameters EXT > 5 arcsec, EXT_STAT > 33, and EXT_DET_STAT > 32, namely the extension, extension likelihood, and detection likelihood, respectively (Pacaud et al. 2006, Faccioli et al. 2018). It has been confirmed from simulations that the C1 class is highly free from contamination by spurious detections or misclassified point-like sources (Pacaud et al. 2007; Clerc et al. 2012).
We updated the current calibration files (CCF) and used an updated (SAS) package (version 15.0.0) rather than the package was used in the preceding steps. That was proceeded by an automated pipeline using python to update and create the required files for the analysis.

### 2.4 Defining galaxy clusters radii and masking other field sources

The circular aperture for each cluster was defined by a semi-interactive procedure developed in (Clerc et al. 2012, Section 2.4). Through that procedure, we determined the radii corresponding to clusters emission extent, namely $R_{fit}$. Firstly, a count rate measurements were performed in three energy bands: [0.5-2], [0.5-1] and [1-2] keV. Count rate is defined as the mean number of cluster photons collected by the three detectors in one second. Some manual redefinition was allowed in the case of adjustment of the cluster position center obtained by XAMIN and accounting for the presence of CCD gap or detector borders in the cluster emission. Secondly, the cluster was assumed to be spherically symmetric and corrected from vignetting and CCD defects. The count rates were measured in concentric annuli using the full pointing exposure to ensure a maximum signal-to-noise ratio (S/N). Another annulus was chosen at a reasonable distance far from the cluster to account for local variation. It was modeled by a photon background and a flat particle background components. The uncertainties on the two parameters of the background and the cluster count rates are derived assuming Poisson noise. Thirdly, we calculated a total count rate by combining each detector count rate measurement and using the total cluster exposure for improving the S/N ratio. Finally, a count rate growth curve as a function of the cluster radius was computed from the total count rate (see Fig. 2), and $R_{fit}$ (the cluster radius) was chosen at the point where the cluster emission merges in the background (background is where S/N ratio equal 1).
The segmentation map is an image with patches represent the pixels ascribed to each source in the observation. It is created by running the SEXTRACTOR software (Bertin & Arnouts 1996) on the wavelet-filtered images (Clerc et al. 2012). It was used in the XAMIN procedure to omit pixels belonging to other sources that obscure the source of interest.
We created a mask out of the segmentation map for each cluster to mask all sources in the field of view but the cluster of interest. The segmentation maps for few cases were edited manually. The patches on the segmentation map might need to be merged together upon visual inspection. There are 11 clusters with modified segmentation maps, we ran all the steps aforementioned to modify the mask created for them accordingly.
The process was fully done by using an automated python pipeline, and then it was checked manually. Images of the masks with the cluster aperture and background annulus overlaying correspond for each observation (see Fig. 3) were created and evaluated by visual inspection. We omitted clusters with background annuli larger than the field of view

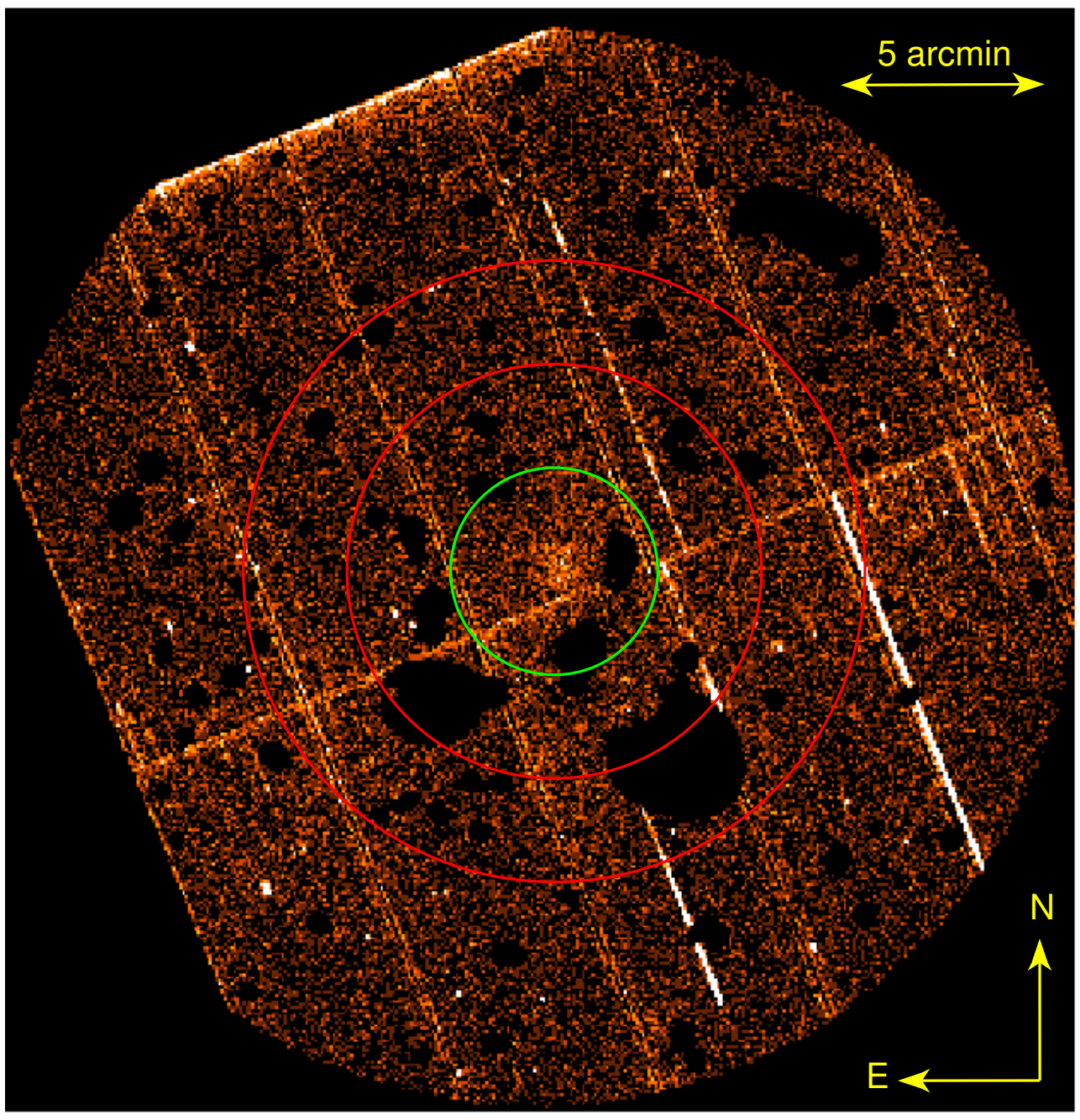

**Figure 3.** The PN image of the cluster X-CLASS 1059. It is overlaid by the cluster extent circle ($R_{fit}$=112.5″) in green and the background annulus in red. The image is multiplied by the segmentation mask, where black colors represent the excluded areas.

(e.g. bright and nearby clusters), more details can be found in Sect. 2.5.2.

### 2.5 Spectral Analysis

The clusters' spectra are extracted from the three EPIC cameras, except one cluster (X-CLASS 2295) that was on the damaged CCD6 of MOS1. The procedure was done by **especget** (SAS task to generate all the necessary files for the spectral fitting of XMM sources) and the fitting is performed in the 0.3-7.0 keV band. We developed a procedure that uses the mask created out of the segmentation map, the cluster radius, and the background annulus in the **especget** to create a spectrum representing a genuine cluster emission with no contamination by nearby sources. The background annulus has an inner radius which is twice the cluster radius ($R_{500}$, $R_{fit}$, $R_{300kpc}$) and the outer radius is triple that value.
For spectral fitting, we utilized the XSPEC package (12.9.1t) (Arnaud 1996) where each cluster spectrum was fitted with a single-temperature APEC (ATOMDB-VERSION 3.0.7) plasma model multiplied by TBABS (Tuebingen-Boulder absorption model by Wilms et al. (2000)) and assuming a fixed galactic hydrogen column density given by Kalberla et al. (2005). The photon counts of each cluster spectrum were grouped into bins with at least one count per bin as it was recommended in (Krumpe et al. 2008) for X-ray spectra, using the Ftools task **grppha**, and we used the Cash statistics (Cash 1979) for fitting the model due to the Poisson nature of the noise.
The spectra for the three cameras were fit simultaneously with the temperature parameters tied together, and all the errors quoted here are with their 1 $\sigma$ errors. The solar abundance was fixed at 0.3 $Z_{\odot}$ and redshifts were described in Sect. 2.2. We calculated the flux and luminosity in the energy band [0.5-2.0] keV. It is not straightforward to calculate the unabsorbed luminosity within its uncertainties while keeping all other parameters at their fitted value, so we used the new convolution model **clumin**. It gives the unabsorbed luminosity and uncertainties for the selected model without changing other parameters (i.e. column density parameter is not equal zero). We determined the bolometric luminosity of the cluster rest frame, in the energy band [0.01-100.0] keV, from a dummy response matrix created by the best fitting model and its parameters. All the errors represent 68% of the confidence range. All the spectra extracted within different apertures are following the same fitting process, and we will describe below the three apertures used in our analysis.

### 2.5.1 $R_{500}$ measurements

$R_{500}$ is defined as the cluster radius embrace mass density at the cluster redshift, which equals 500 times the critical density of the Universe. We used the empirical equation

$$E(z)R_{500} = r3(T_{500}/3keV)^{\alpha/3}), E(z) = (\Omega_M \times (1+z)^3 + \Omega_\Lambda)^{1/2} \quad (1)$$

for the (Tier 1+2+clusters) sample in Sun et al. (2009) to obtain $R_{500}$ value. The values r3=0.600 and $\alpha$ = 1.65 are from (table 6, Sun et al. 2009) and z is the redshift values of our cluster sample.

We started the iteration procedure with an initial temperature value equal to 2.0 keV. First, we calculated $R_{500}$ from equation (1). Secondly, we used the $R_{500}$ obtained from the previous calculation to create new spectra and background files. We do not generate a new response matrix and ancillary response files yet we use existing files from the same cluster. Thirdly, we computed temperature through XSPEC. We used the same spectral analysis approach we followed in Sect. 2.5. Finally, we use the new temperature we obtained from XSPEC to calculate a new $R_{500}$ value using equation (1).

We set the next condition to check the output results and to determine how long the iteration process will proceed for one cluster, also the results for each iteration step are preserved. We start with 5 iterations and create a representative plot for it (see Fig. B1), then we check if any of these iteration changes by less than 0.01 of its precedent. Whether this condition is reliable or not, we run five more iterations and create a representative plot, of all ten iterations, then we check again the previous condition. If that condition is valid, we choose the lowest value (below 0.01) which represent a steady behavior among $R_{500}$ values in the iteration steps and also shows consistency with their uncertainty range in the plot, then the iteration process halt and move to the next cluster. If the previous condition is not true, we run five more iterations and create a representative plot, and again check if the condition valid, then the iteration process halts and move on to a next cluster. If the condition failed then we consider this failed case and move to the next cluster.

Out of the 92 clusters, we managed to calculate $R_{500}$ within the range of 0.5 to 1.24 Mpc for 57 clusters (Table A2). The total counts range between 250 to 2.14×10$^5$ counts with a median value equal 3820 counts. Also, we calculated the bolometric and band [0.5-2.0] keV luminosity for them. The remaining 35 clusters failed because did not fulfill the condition required to pass through the pipeline.

**Table 1.** Categorization of the clusters in each aperture ($R_{500}$, $R_{fit}$, $R_{300kpc}$).

| Aperture | Number of sources | succeed cases | failed cases | literature |
|---|---|---|---|---|
| $R_{500}$ | 92 | 57 | 35 | 02$^{\ddagger}$ |
| $R_{fit}$ | 92 | 68 | 02 | 22 |
| $R_{300kpc}$ | 92 | 87 | 05 | – |

$^{\ddagger}$ They are also presented in the succeed cases with $R_{500}$.

### 2.5.2 $R_{fit}$ measurements

The cluster temperature measured within $R_{fit}$ is called $T_{fit}$, and we were able to derive the temperature for 68 clusters of the cluster sample (Table A3) which have total counts range between 145 to 3.2×10$^4$ counts and a median value equal 2600 counts. Two clusters with unconstrained errors in temperature measurement (X-CLASS 1185, X-CLASS 1813) are removed from the $R_{fit}$ sample. The remaining twenty-two clusters are classified as bright and nearby clusters with a high chance of being studied in the literature, since their circular aperture is fully covering the image and their background annulus extend outside it. It might be surprising that we could not manage to calculate temperature within $R_{fit}$ for the clusters (X-CLASS 458, X-CLASS 2202, X-CLASS 2211) while they are already have temperature measurments within $R_{500}$, but it happened that they reach stable $R_{500}$ lower than $R_{fit}$, so it was possible to obtain their temperature measurements.

We found fourteen of these bright and nearby clusters with studies using XMM-mission, six of them with studies using other X-ray missions, and two with no studies at all (Table A5). For some cases of these 22 clusters, we measured temperatures very close to values found in the literature.

### 2.5.3 $R_{300kpc}$ measurements

We also derived the temperature within a fixed aperture of 300 kpc, and we were able to drive the temperature within this aperture for 87 clusters (Table A4) with total counts in the range of 125 to 2.7×10$^5$ counts and a median value equal 2410 counts. Five clusters (X-CLASS 1307, X-CLASS 1185, X-CLASS 535, X-CLASS 2130, X-CLASS 2209) are removed from the $R_{300kpc}$ sample because they have unconstrained errors on their temperature measurements. It is worth noting that for the cluster (X-CLASS 1307) we managed to derive its temperature using $R_{fit}$. Also, we found that using the chi-squared (chi) fit statistic instead of the Cash statistics gave acceptable temperature for that cluster (X-CLASS 1307) within aperture of 300 kpc.

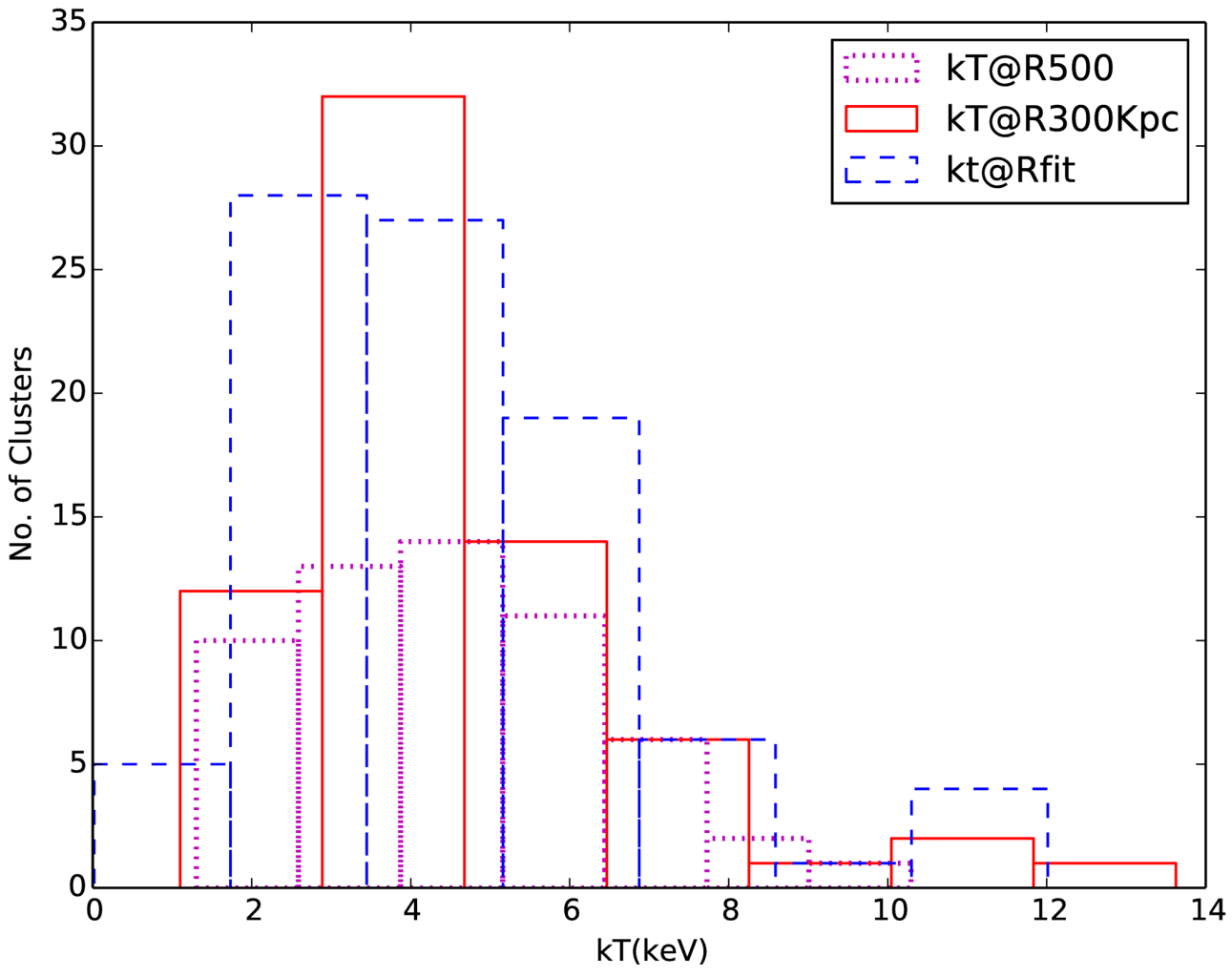

**Figure 4.** Temperature distribution of the galaxy clusters in XC1-RM. The temperature derived within $R_{fit}$ is represented by blue dashed line (68 clusters), within $R_{300kpc}$ is represented by red solid line (87 clusters), and within $R_{500}$ is represented by pink dotted line (57 clusters). The three groups almost peaks at the same range of temperatures.

## 3 RESULTS

### 3.1 Temperature measurements

We discussed in Sect. 2.5 the procedure which we followed to calculate temperature within different apertures and the summary of the results is in Table 1.
Our cluster sample span temperature range of (1.0-10 keV) within $R_{500}$ with a peak at nearly 4.0 keV and a mean value ~ 4.5 keV in the three apertures ($R_{500}$, $R_{fit}$, $R_{300kpc}$). The temperature distribution within $R_{500}$, $R_{fit}$, and $R_{300kpc}$ are shown in Fig. 4. The mean luminosity in the band [0.5-2.0] keV for our sample is $12{\times}10^{43}$ erg $s^{-1}$, $8{\times}10^{43}$ erg $s^{-1}$, and 6 $\times10^{43}$ erg $s^{-1}$ within $R_{500}$, $R_{fit}$, and $R_{300kpc}$, respectively. While the mean bolometric luminosity for our sample are $4{\times}10^{44}$ erg $s^{-1}$, $2{\times}10^{44}$ erg $s^{-1}$, and $2{\times}10^{44}$ erg $s^{-1}$ within $R_{500}$, $R_{fit}$, and $R_{300kpc}$, respectively.
Table A2, Table A3, and Table A4 represent the clusters X-ray properties measured within $R_{500}$ (57 clusters), $R_{fit}$ (70 clusters), and $R_{300kpc}$ (92 clusters), respectively. In Table A2; we present the characteristic properties of each cluster; X-CLASS ID, is presented in column [1]. Temperature and its negative and positive 68% uncertainty are presented in columns [2], [3], and [4], respectively. $R_{500}$ aperture is in column [5], and fluxes measured in band [0.5-2 keV] are in column [8]. The band luminosity [0.5-2 keV] is presented in column [11]. In Table A3; the first column is the X-CLASS ID, where $R_{fit}$ is in column [2]. Temperature and its negative and positive 68% uncertainty are presented in columns [3], [4], and [5], respectively. Fluxes measured in band [0.5-2 keV] are in column [6]. The band luminosity [0.5-2 keV] is listed in column [9]. In Table A4; The first column is the same like Tables (A2, A3), $R_{300kpc}$ measured in arcses in column [2], and temperature and its negative and positive 68% uncertainty are given in columns [3], [4], and [5], respectively. Fluxes and the luminosity measured in band [0.5-2 keV] are in column [6], and [9], respectively.

For each of the twenty-two bright clusters in Sect.2.5.2, we searched for temperature measurement results in the literature which should fulfill the following conditions: the cluster core should not be excluded, it should be XMM study, and the temperature measured in circular aperture not annuli. Table A5 lists the results we found in the literature and their references. In Table A5, we also present our results which we managed to calculate for some clusters. The first column represents the X-CLASS ID, the temperatures we found in the literature with its uncertainty are presented in the sub-columns [1], [2], [3], and [4], respectively. The aperture where those published temperatures are measured and the references are presented in the sub-columns [5], and [6], respectively. The second column represents our results and it presents our temperatures and its negative and positive 68% uncertainty in the sub-columns [1], [2], and [3], respectively. $R_{500}$ is presented in sub-column [4].

### 3.2 L-T Relation

We investigated the band [0.5-2 keV] luminosity-temperature (L-T) relation for the clusters having $R_{500}$ values. They are 57 clusters (Table 1, Sect. 2.5.1). Figure 5 shows the L-T relation for our 57 clusters along with 1 $\sigma$ uncertainty. The relation was fitted by the power law equation $[L/L_0 = E(z)^n \times b \times (T/T_0)^a]$ in log space of base 10, where n = 1 assuming self-similar evolution, and $b$, $a$ are the normalization and the slope, respectively. We used the orthogonal regression model (BCES) to fit the power law equation to the sample and assumed that $L_0 = 5\times10^{43}$ erg $s^{-1}$, and $T_0 = 4$ keV. The solid red line represents the power law fit of our cluster sample. We found a slope $a$ =3.00 ± 0.31, and normalization $b$ = 1.07 ± 0.12. Subsequently, we compared our results with REXCESS (Pratt et al. 2009, hereafter P09), XMM-LSS (Clerc et al. 2014, hereafter C14) and XXL (Giles et al. 2016, hereafter XXL) samples of clusters. We used our analytic expression for the L-T relation to fit their data using the same pivot $L_0$ and $T_0$ of our relation. The green dashed line represents the fit to XXL sample which gives slope $a$ =3.03 ± 0.26, and normalization $b$ = 1.29 ± 0.09, where the blue dash-dotted line represents the fit to P09 sample which gives slope $a$ =2.97 ± 0.31, and normalization $b$ = 2.19 ± 0.50. The purple solid line represents the fit to C14 sample which gives slope $a$ =3.10 ± 0.38, and normalization $b$ = 1.82 ± 0.44. The intrinsic logarithmic scatter of our L-T relation is calculated following the procedure presented by Arnaud et al. (2005). All the fitting parameters along with the intrinsic scatter are listed in Table 2.

#### 3.2.1 $L_{bol}$-T Relation

Since we calculated the bolometric luminosity for the 57 clusters having $R_{500}$ values (Sect. 2.5.1), we investigated their bolometric luminosity-temperature ($L_{bol}$-T) relation. The sample data was fitted using the same power law equation used in Sect. 3.2. We found a slope = 3.25 ± 0.24, normalization = 2.63 ± 0.30 and intrinsic scatter= 0.38 ± 0.5 comparable to the results in Zou et al. (2016) and P09. Zou et al. (2016) studied a sample of 23 clusters and their core-included (L-T) relation gave a slope = 3.28 ± 0.33 and P09

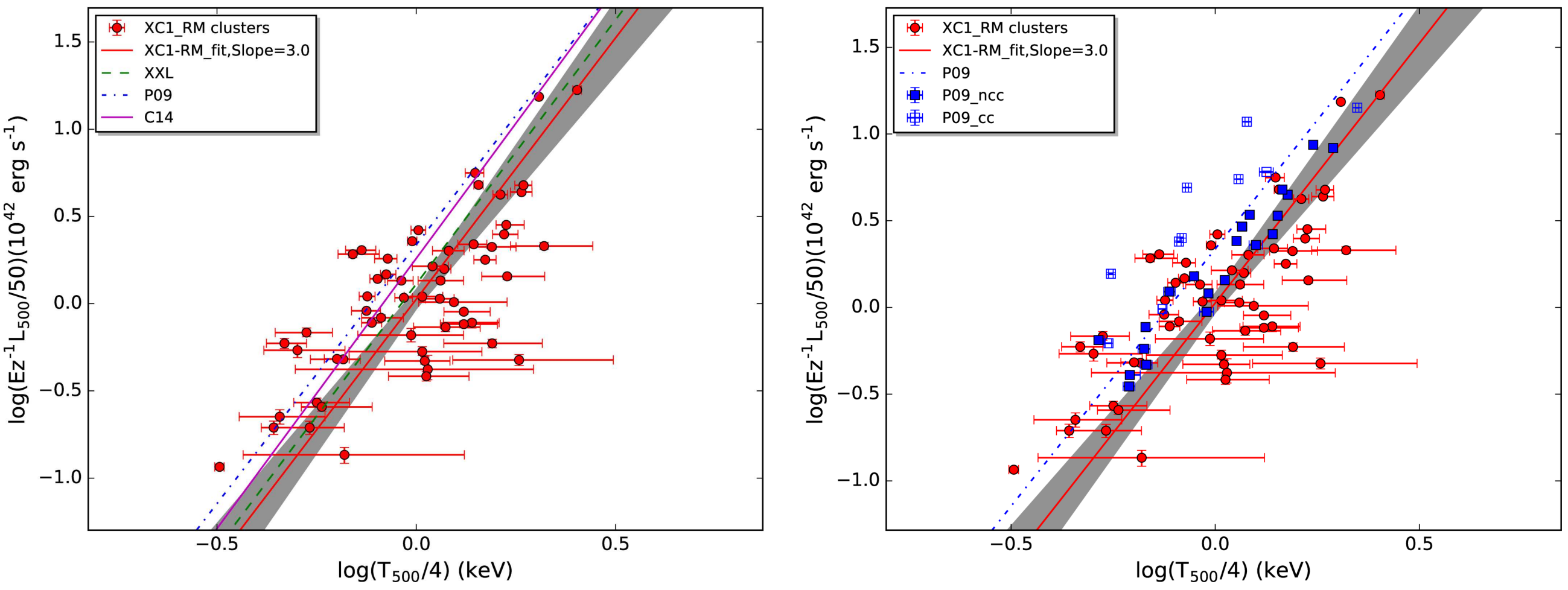

**Figure 5.** L-T relation for the 57 clusters. Left: The red solid line represents the fit to our cluster sample in red points. The shaded area represents the 1 $\sigma$ uncertainty for our relation. The magenta solid line, green dashed line, and the blue dash-dotted line represent the fit to Clerc et al. (2014), Giles et al. (2016) and Pratt et al. (2009), respectively. Right: The same as the left plot, but it shows P09 sample divided into cool-core (empty blue square) and non cool-core (filled blue square).

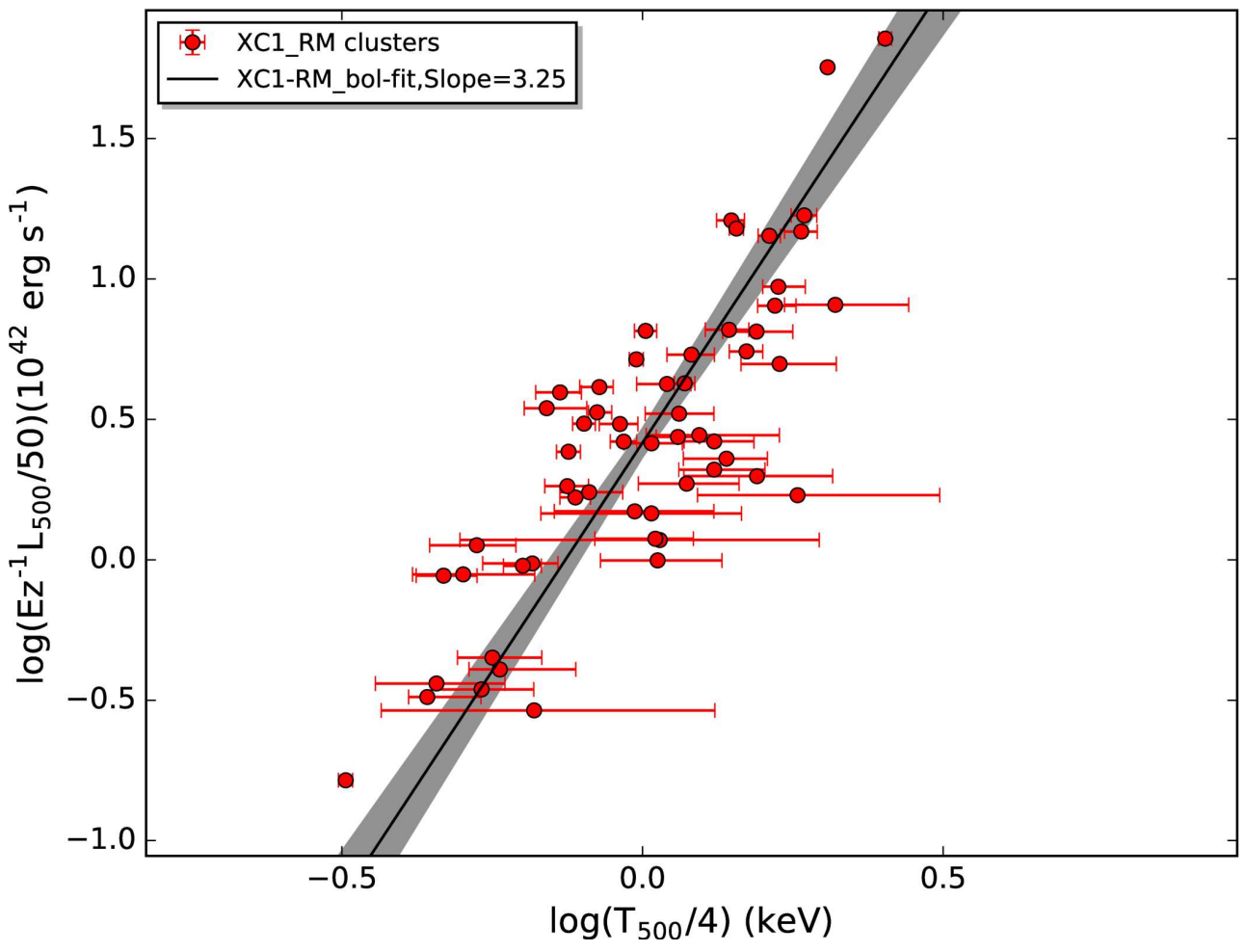

**Figure 6.** The $L_{bol}$-T relation. The black solid line represents the fit to our sample in red points. The shaded area represent the 1 $\sigma$ uncertainty for our sample.

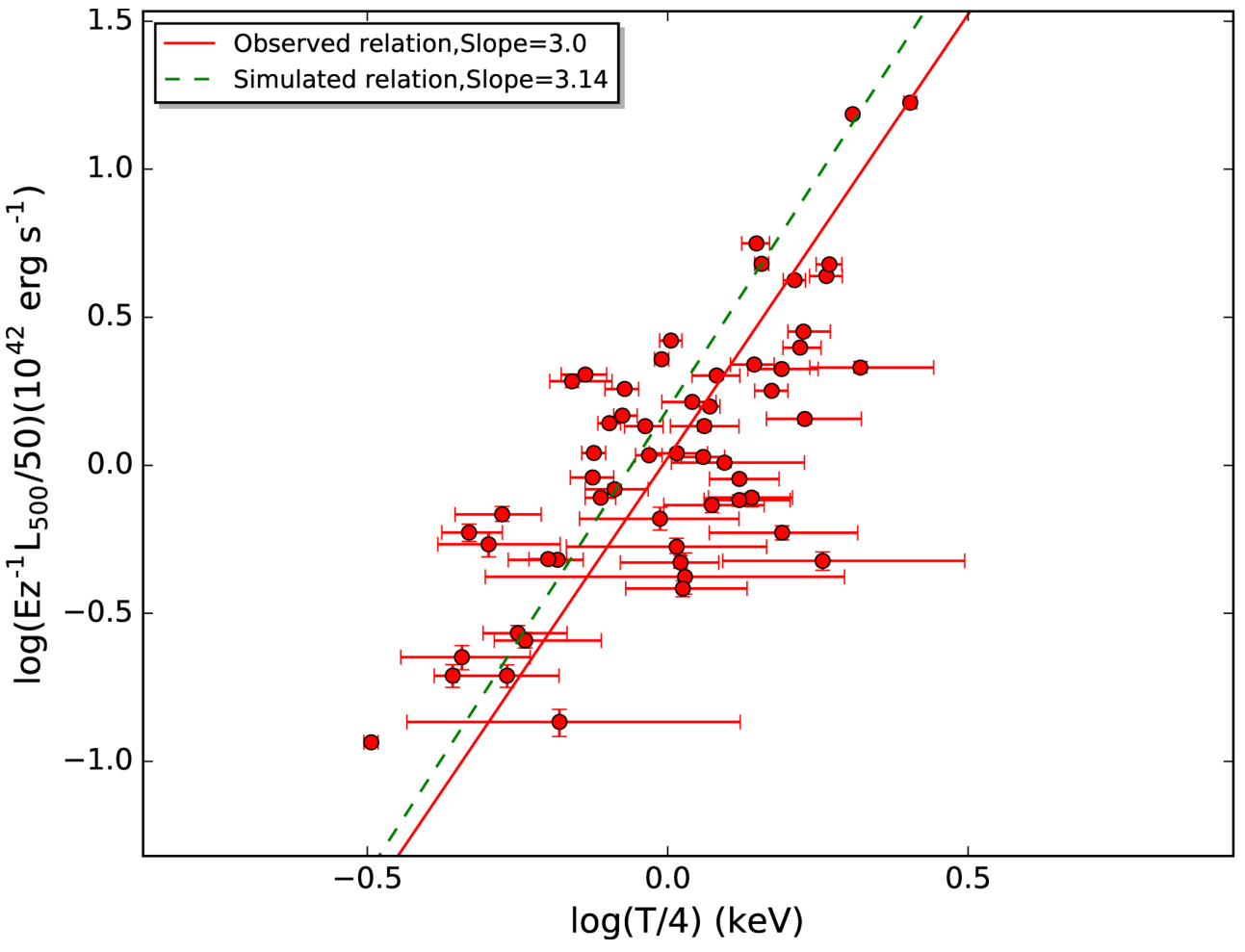

**Figure 7.** The observed and simulated L-T relation. The red solid line represents the fit to our sample in red points. The dashed green is the simulated L-T relation

.

found a slope = 3.35 ± 0.32 using also bolometric luminosity for the core included sample. Figure 6 shows the $L_{bol}$-T relation for our sample with 1 $\sigma$ uncertainty and the fit results are given in Table 2.

### 3.3 Simulated L-T relation

We want to investigate the effect of Malmquist bias on our sample, where it would present bright clusters in our sample than the true distribution at a given redshift (Clerc et al. 2014; Mantz et al. 2010b; Pacaud et al. 2007). Thus, we developed a simulation approach to estimate the amount of bias affecting our observed L-T relation. We created a sample of points that represents the underlying population of clusters and considered our observed band L-T relation as the unbiased relation to obtain a realization of the sample luminosities. Then, we applied detection constrains (e.g. flux threshold, background noise, and exposure time). The simulated sample spectral properties and their uncertainties are introduced like our observed sample, and we used the BCES fit to investigate the resulted L-T relation. We describe below the simulation approach which we followed in details.
We created a list of 10000 random T and z values (sample 1) distributed according to a galaxy cluster mass function (Tinker et al. 2008), WMAP5 cosmological parameters, and the M-T relation from Arnaud et al. (2005). A second list of 500000 random T and z values (sample 2) was created using the same mass function, WMAP9 cosmological parameters, and a different M-T relation from Sun et al. (2009).
We used our L-T relation parameters and the power law equation in Sect.3.2, $L/L_0 = E(z)^n b(T/T_0)^a$ with $b$=1.07, $a$=3.0, n = 1, intrinsic scatter $\sigma_{int,L}$ = 0.44, $L_0 = 5 \times 10^{43}$ erg s$^{-1}$, and $T_0$ = 4 keV, to calculate the luminosity (L) for the temperature values in the two samples.

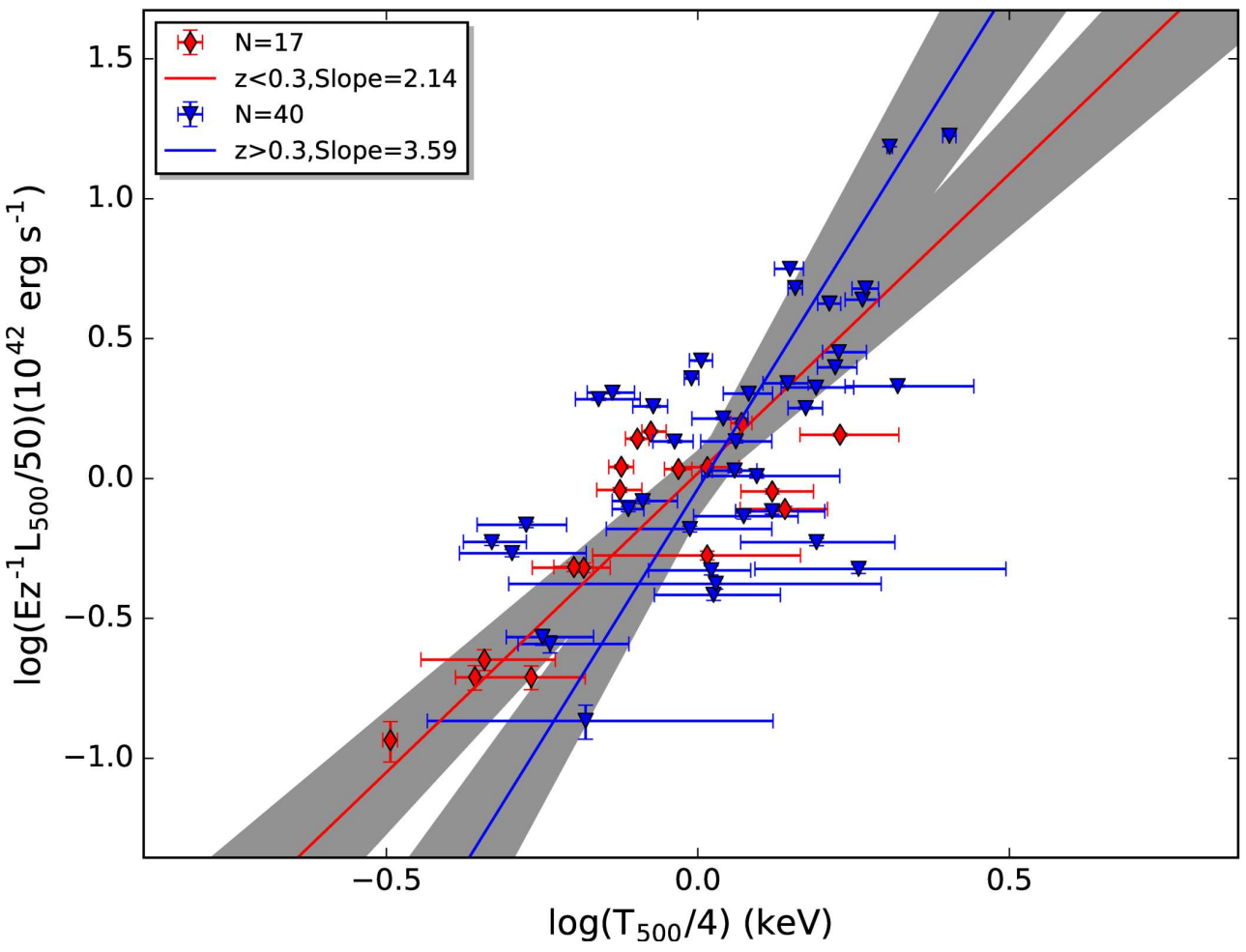

**Figure 8.** L-T relation in different redshift bins. The red points represent clusters with redshift lower than 0.3. The blue points represent clusters with redshift higher than 0.3. The shaded area for each of the two samples represented the 1 $\sigma$ uncertainty.

We utilized the same XSPEC package (12.9.1t) to calculate the flux for both samples. We used a single-temperature APEC plasma model multiplied by TBABS absorption model (Wilms et al. 2000) and assuming a fixed galactic hydrogen density column at $0.026\times10^{22}$ cm$^{-2}$. The solar abundance was fixed at 0.3 $Z_\odot$ and the redshift was obtained from the two samples. The flux was calculated by changing the normalization parameter iteratively to give luminosity values nearly equal to the ones calculated from the L-T relation. The flux limits of our observed sample is above $2.48\times10^{-14}$ erg s$^{-1}$ cm$^{-2}$ and below $1.5\times10^{-12}$ erg s$^{-1}$ cm$^{-2}$. We applied the same flux limit to the simulated samples which reduced the number of the two samples to around 300 simulated points in sample 1 and 12000 simulated points in sample 2. After that, the **fakeit** command in the XSPEC package was used to create spectra of the parameters obtained in the preceding steps. The rmf and arf in (EPIC) thin filter used in the analysis are from the XMM online database ¶. Then, a random background file from our sample (to simulate the noise we found in the real observation) was applied, and we chose an average exposure time of 15 ks (average exposure time for our sample). At that point, we created mock spectra for both samples (1 and 2). To derive temperatures and luminosities we followed the same procedure mentioned in Sect. 2.5. For both samples, we generated a random subsamples, where each contains 57 simulated points, to resemble the observed cluster sample. Subsequently, we plotted the luminosity versus temperature (L-T) for the subsamples and the orthogonal regression model (BCES) was applied for fitting (Sect.3.2). To determine an average slope and intercept, we calculated the mean for 1000 subsamples in both samples. We found a slope of a =3.18 ± 0.07, and normalization b = 1.58 ± 0.22 for sample 1 and a slope of a =3.14 ± 0.27, and normalization b = 1.58 ± 0.23 for sample 2. The result of sample 2 are shown in Fig. 7 and Table 2.

**Table 2.** The L-T relation parameters, $L/L_0 = E(z)^n b(T/T_0)^a$, where $L_0 = 5\times10^{43}$ erg s$^{-1}$, and $T_0$ = 4 keV. We used the BCES orthogonal fit method. We use the first subsample in Table 1 for all our L-T relations

| clusters sample | a | b | $\sigma_{int,L}$ |
|---|---|---|---|
| XC1-RM | 3.00 ± 0.31 | 1.07 ± 0.12 | 0.44 ± 0.06 |
| XC1-RM$_{\rm bol}$ | 3.25 ± 0.24 | 2.63 ± 0.30 | 0.37 ± 0.05 |
| XC1-RM$_{\rm Simulated}$ | 3.14 ± 0.27 | 1.58 ± 0.23 | 0.38 ± 0.05 |
| XC1-RM $_{z<0.3}$ | 2.14 ± 0.28 | 1.05 ± 0.19 | 0.18 ± 0.04 |
| XC1-RM $_{z>0.3}$ | 3.59 ± 0.54 | 0.91 ± 0.19 | 0.42 ± 0.07 |
| XXL | 3.03 ± 0.26 | 1.29 ± 0.09 | 0.40 ± 0.04 |
| P09 | 2.97 ± 0.31 | 2.19 ± 0.50 | 0.34 ± 0.06 |
| C14 | 3.10 ± 0.38 | 1.82 ± 0.44 | 0.37 ± 0.05 |

## 4 DISCUSSION

In this section, we interpret our temperature measurements and luminosity-temperature relation results and compare them with other related studies in the literature. The effect of the Malmquist bias on the observed L-T relation is also discussed.

### 4.1 Temperature measurements

In Fig. 9, we compare our results with 16 and 20 galaxy clusters from XCS-DR1 and 2XMMi/SDSS catalogues matched within 15″, respectively.

XCS-DR1 is the XMM Cluster Survey first data release (Hilton et al. 2012; Mehrtens et al. 2012; Lloyd-Davies et al. 2011). Hilton et al. (2012) investigated the evolution of the L-T relation for 211 clusters of XCS-DR1 with spectroscopic redshifts, where the temperature and luminosity were measured using the pipeline described in Lloyd-Davies et al. (2011). For XCS temperature measurements, our results are nearly consistent (left panel, Fig. 9). We could not compare our $R_{\rm fit}$ with their radii measurements, but it is possible that different cluster emission aperture resulted in slightly different temperature values.

Takey et al. (2013) presented 345 galaxy clusters with temperature measured within the optimum extraction radius and bolometric luminosity measured within $R_{500}$. Its optimum extraction radius described as the radius representing the highest signal-to-noise ratio (S/N) point on a radial profile of the X-ray surface brightness of each cluster created in the energy band [0.5-2.0] keV on each camera (PN, MOS1, MOS2) and the combined one (Sect.3.2.1, Takey et al. 2011). For 2XMMi/SDSS temperature measurements, our results are nearly consistent (right panel, Fig. 9) except few points that need more investigation. Thus, we compared the radii used in both samples, which were originated from different techniques (Sect.3.2.1, Takey et al. 2011), (Sect.4, Clerc et al. 2014). Interestingly, we found that all radii extracted in 2XMMi/SDSS are lower than the radii in our sample. Nonetheless, the temperature measurement for the clusters which have the highest difference in radii measurements are not affected (see inset, right panel, Fig. 9). This shows that using different clusters extent is not always the main reason leading to discrepancy in temperature measurements, and such discrepancy might rise from different reasons (e.g., spectral fitting models).

¶ ftp://legacy.gsfc.nasa.gov/xmm/data/responses/qd/

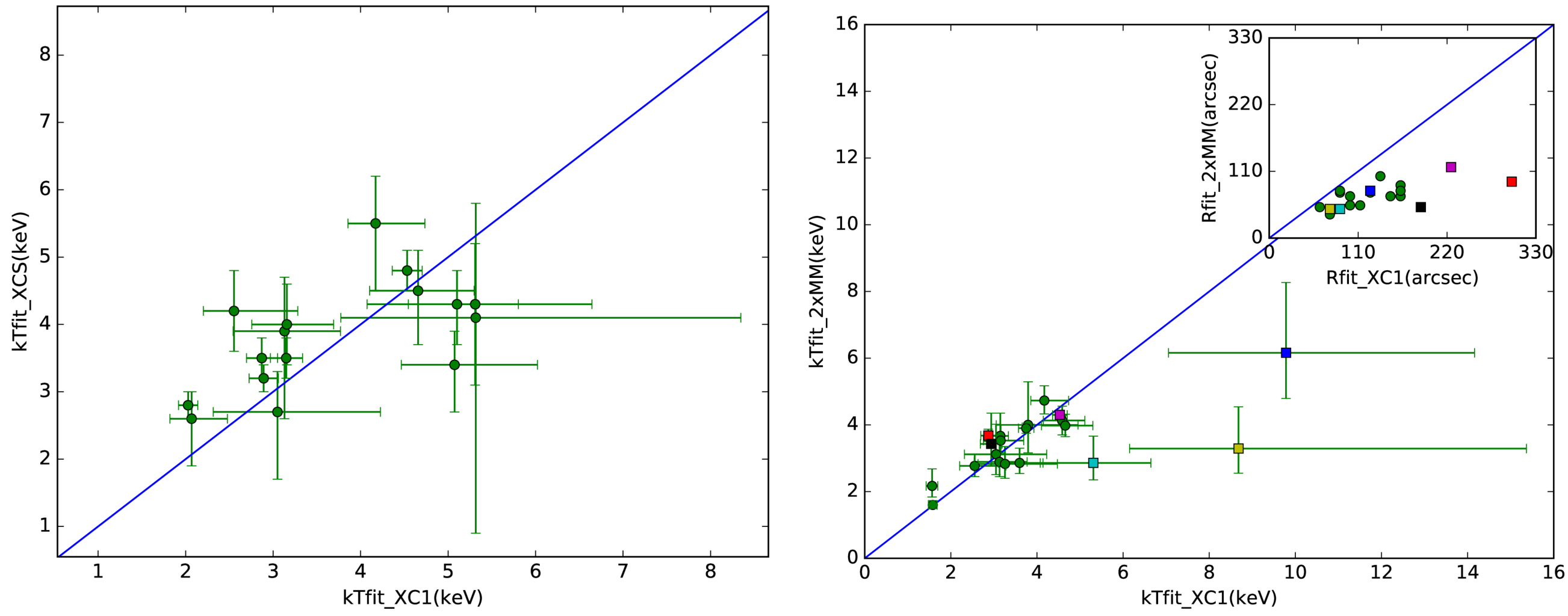

**Figure 9.** Temperature measurements comparison with XCS and 2XMMi/SDSS. Left: A comparison between the temperature of 16 galaxy clusters in common between XC1-RM and XCS (Hilton et al. 2012). The solid blue line shows the 1-1 relationship. Right: A comparison between the temperature of 20 galaxy clusters in common between XC1-RM and 2XMMi/SDSS (Takey et al. 2013). The solid blue line shows the 1-1 relationship, and the inset is a comparison between the radii measurements of the two samples. The color points represent the outliers on the inset plot.The errors represent 68% of the confidence range.

Our temperature results are mostly consistent with the results in XCS-DR1 and 2XMMi/SDSS, where the existence of a few clusters having high error bars in our sample is prominent since we do not exclude clusters with high uncertainties similar to XCS and 2XMMi/SDSS study. Additionally, XCS-DR1 and 2XMMi/SDSS are using different spectral fitting models and different methods to determine the cluster emission extent.

Our clusters sample has an average temperature of 4.0 keV in the three sub-classes extracted in different radii. Similar result can be found in Giles et al. (2016) and Ebrahimpour et al. (2018).

### 4.2 L-T relation

We chose the BCES method (Akritas & Bershady 1996) because it has been widely used in other studies to which we will compare our results. The errors in temperature and luminosity were converted into a log-normal likelihood using the method of Andreon (2012). We preferred to use one BCES fitting method (BCES orthogonal) which minimizes the orthogonal distance from the data points to the fitted line.

We found a steeper slope than that was found in the self-similar expectation (~2.0), which is in an agreement with recent studies (Lovisari et al. 2015; Takey et al. 2019; Hilton et al. 2012; Maughan et al. 2012; Giles et al. 2016; Pratt et al. 2009). It has been found that cluster samples constructed without regard to their cores activity like our current sample almost give a slope $\geq 3$ (Giles et al. 2016; Clerc et al. 2014; Mantz et al. 2010b). Our observed L-T relation gave a $=3.00 \pm 0.31$, and b $= 1.07 \pm 0.12$ which is consistent with XXL results (a $=3.03 \pm 0.26$, b $= 1.29 \pm 0.09$) within $1\sigma$ in normalization.

Comparing with P09, we found a normalization within $2\sigma$ with their results which was led by the strong cool-core present in their sample (Right plot in Fig. 5). P09 sample has high luminosity clusters at redshifts z< 0.2 and flux limit above $3 \times 10^{-12}$ erg s$^{-1}$ cm$^{-2}$ (Böhringer et al. 2007) opposite to our sample which contains low luminous deep clusters (flux below $1.5 \times 10^{-12}$ erg s$^{-1}$ cm$^{-2}$, redshifts up to 0.6). We do not know how many cool-core clusters are in our sample, since we could not excise them.

In Fig. 8 we divide our sample in two redshift bins to investigate the L-T parameters evolution with redshift. We found that the slope is $3.59 \pm 0.54$ at redshift greater than 0.3, which is almost as twice as the slope $2.14 \pm 0.28$ at redshift less than 0.3, and we found a slight change in normalization within its error range, (See Table 2).

Our result does not agree with XCS-DR1 sample (Hilton et al. 2012), where they found that the slope did not change dramatically with redshift. In XCS-DR1, they divided the sample into three redshift bins and they fitted them using orthogonal method. Comparing with their first two redshift bins subsamples, which have a slope equals $3.18 \pm 0.22$ in the redshift bin lower than 0.25, and a slope equals $2.82 \pm 0.25$ in the redshift bin higher than 0.25, they found that the slope decreased as the redshift increased contrary to our findings.

On the other hand, in 2XMMi/SDSS (Takey et al. 2013), they found a slope equals $2.55 \pm 0.23$ in the redshift bin below 0.25 and a slope equals $3.27 \pm 0.26$ in the redshift bin above 0.25. The slope is higher in high redshift bins, which is in accordance with our results, but they suggested that this is resulted from including clusters and groups of low temperature in the low redshift bin.

To investigate this result further, we used a different redshift thresholds (e.g.: z greater than 0.25, and z less than 0.25),

where we found that the slope still higher at the higher redshift bins than in the lower redshift bins.
Also, we found that the low temperature clusters in our sample are presented in the two redshift bins, so it was not the cause of shallower slope in the low redshift bins. We suspect that the presence of high luminosity and intermediate temperature clusters are the reason of a higher slope in higher redshift bins.

### 4.3 Effect of the Malmquist bias

We have shown that using a developed approach (Sect. 3.3) to investigate the Malmquist bias on the L-T relation produce a steeper slope than the one we found in the observed-uncorrected L-T relation (Fig 7). It gave a slope and normalization ( a= 3.14 ± 0.27, b= 1.58 ± 0.50) higher than the input parameter (a=3.00, b= 1.07). Consequently, it appears that the Malmquist bias raised our L-T relation parameters which means there is a possibility to find a shallower slope, if we undertook a sophisticated approach (accounting for the biases) to correct our observed-uncorrected L-T relation.
Although we can not confirm the concluded result, or make a proper comparison with other studies, but we found comparable results. In XXL (Giles et al. 2016), the selection function of the survey was taken into account where they found a shallower slope after applying the correction to the L-T relation. Also in Bharadwaj et al. (2015), they found a shallower slope for the HIFLUGCS galaxy clusters and its sub-samples after correcting the sample biases (Table 3, Bharadwaj et al. 2015). It is worth noting that in Bharadwaj et al. (2015), the results were for a corrected bolometric L-T relation and that they used BCES L|T method.
It is also worth mentioning that in Zou et al. (2016), their slope did not vary significantly after applying similar bias correction. This was attributed to the BCES regression method used, since they used BCES orthogonal over BCES L|T, where the later minimizes the residuals in L, so it is more sensitive to the selection effects.
We also investigated the probability that the inputs of the simulation approach might have a major impact on the output results. In Sect. 3.3 we used a different set of cosmological parameter and M-T relations to investigate their impact on the simulation approach. We found that changing the M-T relation nor the cosmological parameters has a significant impact on the simulated L-T relation. Furthermore, we applied a different L-T relation in the simulation approach, we used the XXL corrected L-T relation (a= 2.63, b= 0.71, $\sigma$ = 0.47) and it gave a higher slope = 2.98 and normalization=0.97. That investigation and the aforementioned result that we obtained at different redshift bins (Sect. 4.2) could present reasonable view on the bias affect our sample.

## 5 SUMMARY

In this work, we introduced the first X-ray study of (X-CLASS-redMaPPer) sample in the X-CLASS survey, which contains 92 galaxy clusters. The sample spans a range of redshifts of $0.05 < z < 0.6$. Here, we summarize our main results:

(i) Our spectral results are in a good agreement with the results in literature. We measured the X-ray spectral properties for the 92 galaxy clusters and succeed to obtain the results for 57, 68, and 87 clusters within $R_{500}$, $R_{\rm fit}$, and $R_{\rm 300kpc}$, respectively. The X-ray spectral properties were measured in a homogeneous way by an automated pipeline that we developed for this work.
(ii) For the measurements within the $R_{500}$ aperture, the sample temperature spans range from T ≃ 1.0 keV to 10 keV and the band luminosity[0.5-2.0] keV ranges from L ≃ $6{\times}10^{42}$ to $11{\times}10^{44}$ erg s$^{-1}$. $R_{500}$ was calculated in an automated iterative method, and the procedure also created representative plots for manual checks.
(iii) We studied the X-ray luminosity-temperature (L-T) relation in log-log space for the clusters parameters measured within $R_{500}$, which gave a slope =3.00 ± 0.31 supporting studies that have a higher slope than the one expected by self-similar model, and in a good agreement with similiar recent studies.
(iv) We divided the sample into two redshift bins, where we found a steeper slope in the high redshift bin. The slope is always steeper in higher redshift bins than in lower redshift bins opposite to what is found in literature, indicating that the slope in higher redshift bins could be biased by the existence of high luminosity clusters, and we will investigate this in a larger cluster sample to confirm this behaviour.
(v) We developed a simplified simulation approach to asses the amplitude of bias affecting our L-T relation, where we found that the bias in our sample is moving the L-T relation parameters higher. There are recent studies that found such result, but it requires further investigation.

Future investigations will determine the intracluster gas mass ($M_{\rm gas}$) of the sample. We will probe its scaling with other X-ray quantities. The impact of selection effects on the scaling relations will be quantitatively assessed by increasing the size of the sample and implementing elaborated approaches taking advantage of the enhanced statistics and redshift leverages.

## ACKNOWLEDGEMENTS

This work was done in Egypt and supported by the National Research Institute of Astronomy and Geophysics (NRIAG) through funds for short international travels. The results presented here are based on data from the X-CLASS serendipitous cluster catalogue extracted from the XMM archival data. I am thankful to FACCIOLI Lorenzo for his help on the data sample. Mona.M acknowledges support from IAU via air flight tickets for a short visit to The Inter-University Centre for Astronomy and Astrophysics (IUCAA). I also thank the (IUCAA) institute for accepting my visit and hosting for a few months. I am thankful to Prof. Ajit Kembhavi and Prof. Ranjeev Misra for their helping and guiding during the visit. My thanks to Anjlia who helped me on the **fakeit** procedure in XSPEC. This research has made use of software provided by the High Energy Astrophysics Science Archive Research Center (HEASARC), which is a service of the Astrophysics Science Division at NASA/GSFC and the High Energy Astrophysics Division of the Smithsonian Astrophysical Observatory. Funding for the Sloan Digital Sky Survey (SDSS) has been provided by the Alfred P. Sloan Foundation, the Participating Institutions, the National Aeronautics and Space Administration, the National Science Foundation, the U.S. Department of Energy, the Japanese Monbukagakusho, and the Max Planck Society. The SDSS Web site is http://www.sdss.org/. The SDSS is managed by the Astrophysical Research Consortium (ARC) for the Participating Institutions. The Participating Institutions are The University of Chicago, Fermilab, the Institute for Advanced Study, the Japan Participation Group, The Johns Hopkins University, Los Alamos National Laboratory, the Max-Planck-Institute for Astronomy (MPIA), the Max-Planck-Institute for Astrophysics (MPA), New Mexico State University, University of Pittsburgh, Princeton University, the United States Naval Observatory, and the University of Washington. Also, We thank The Kottamia Center of scientific Excellence.

## REFERENCES

Abolfathi B., et al., 2018, ApJS, 235, 42
Akritas M. G., Bershady M. A., 1996, ApJ, 470, 706
Allen S. W., Evrard A. E., Mantz A. B., 2011, ARA&A, 49, 409
Andersson K., Peterson J. R., Madejski G., Goobar A., 2009, ApJ, 696, 1029
Andreon S., 2012, A&A, 546, A6
Arnaud K. A., 1996, in Jacoby G. H., Barnes J., eds, Astronomical Society of the Pacific Conference Series Vol. 101, Astronomical Data Analysis Software and Systems V. p. 17
Arnaud M., Pointecouteau E., Pratt G. W., 2005, A&A, 441, 893
Bertin E., Arnouts S., 1996, A&AS, 117, 393
Bharadwaj V., Reiprich T. H., Lovisari L., Eckmiller H. J., 2015, A&A, 573, A75
Böhringer H., et al., 2000, The Astrophysical Journal Supplement Series, 129, 435
Böhringer H., et al., 2004, A&A, 425, 367
Böhringer H., et al., 2007, Astronomy and Astrophysics, 469, 363
Cash W., 1979, ApJ, 228, 939
Clerc N., Sadibekova T., Pierre M., Pacaud F., Le Fèvre J.-P., Adami C., Altieri B., Valtchanov I., 2012, MNRAS, 423, 3561
Clerc N., et al., 2014, MNRAS, 444, 2723
Croston J. H., et al., 2008, A&A, 487, 431
Ebrahimpour L., et al., 2018, arXiv e-prints, p. arXiv:1805.03465
Faccioli L., et al., 2018, A&A, 620, A9
Finoguenov A., et al., 2007, ApJS, 172, 182
Giles P. A., et al., 2016, A&A, 592, A3
Gioia I. M., Maccacaro T., Schild R. E., Wolter A., Stocke J. T., Morris S. L., Henry J. P., 1990, ApJS, 72, 567
Hilton M., et al., 2012, MNRAS, 424, 2086
Horner D. J., Perlman E. S., Ebeling H., Jones L. R., Scharf C. A., Wegner G., Malkan M., Maughan B., 2008, ApJS, 176, 374
Kaiser N., 1986, MNRAS, 222, 323
Kalberla P. M. W., Burton W. B., Hartmann D., Arnal E. M., Bajaja E., Morras R., Pöppel W. G. L., 2005, A&A, 440, 775
Kotov O., Vikhlinin A., 2005, ApJ, 633, 781
Krumpe M., et al., 2008, A&A, 483, 415
Lloyd-Davies E. J., et al., 2011, MNRAS, 418, 14
Lovisari L., Reiprich T. H., Schellenberger G., 2015, A&A, 573, A118
Mantz A., Allen S. W., Rapetti D., Ebeling H., 2010a, MNRAS, 406, 1759
Mantz A., Allen S. W., Ebeling H., Rapetti D., Drlica-Wagner A., 2010b, MNRAS, 406, 1773
Maughan B. J., Jones C., Forman W., Van Speybroeck L., 2008, ApJS, 174, 117
Maughan B. J., Giles P. A., Randall S. W., Jones C., Forman W. R., 2012, MNRAS, 421, 1583
Mehrtens N., et al., 2012, MNRAS, 423, 1024
Mittal R., Hicks A., Reiprich T. H., Jaritz V., 2011, A&A, 532, A133
Pacaud F., et al., 2006, MNRAS, 372, 578
Pacaud F., et al., 2007, MNRAS, 382, 1289
Pacaud F., et al., 2016, A&A, 592, A2
Pierre M., et al., 2004, Journal of Cosmology and Astro-Particle Physics, 2004, 011
Pierre M., et al., 2016, A&A, 592, A1
Planck Collaboration et al., 2011, A&A, 536, A11
Pratt G. W., Arnaud M., 2003, A&A, 408, 1
Pratt G. W., Arnaud M., 2005, A&A, 429, 791
Pratt G. W., Croston J. H., Arnaud M., Böhringer H., 2009, A&A, 498, 361
Reiprich T. H., Böhringer H., 2002, ApJ, 567, 716
Rozo E., et al., 2010, ApJ, 708, 645
Rykoff E. S., et al., 2014, ApJ, 785, 104
Sadibekova T., Pierre M., Clerc N., Faccioli L., Gastaud R., Le Fevre J.-P., Rozo E., Rykoff E., 2014, A&A, 571, A87
Sehgal N., et al., 2011, ApJ, 732, 44
Sun M., Voit G. M., Donahue M., Jones C., Forman W., Vikhlinin A., 2009, ApJ, 693, 1142
Takey A., Schwope A., Lamer G., 2011, A&A, 534, A120
Takey A., Schwope A., Lamer G., 2013, A&A, 558, A75
Takey A., Schwope A., Lamer G., 2014, A&A, 564, A54
Takey A., Durret F., Mahmoud E., Ali G. B., 2016, A&A, 594, A32
Takey A., Durret F., Márquez I., Ellien A., Molham M., Plat A., 2019, MNRAS, 486, 4863
Tinker J., Kravtsov A. V., Klypin A., Abazajian K., Warren M., Yepes G., Gottlöber S., Holz D. E., 2008, ApJ, 688, 709
Vikhlinin A., McNamara B. R., Forman W., Jones C., Quintana H., Hornstrup A., 1998, ApJ, 502, 558
Vikhlinin A., et al., 2009a, ApJ, 692, 1033
Vikhlinin A., et al., 2009b, ApJ, 692, 1060
Voit G. M., Kay S. T., Bryan G. L., 2005, MNRAS, 364, 909
Wen Z. L., Han J. L., Liu F. S., 2009, ApJS, 183, 197
Wilms J., Allen A., McCray R., 2000, ApJ, 542, 914
Zou S., Maughan B. J., Giles P. A., Vikhlinin A., Pacaud F., Burenin R., Hornstrup A., 2016, MNRAS, 463, 820
Šuhada R., et al., 2012, A&A, 537, A39

## APPENDIX A: TABLES

Table A1: The X-CLASS-redMaPPer 92 clusters sample.

| *X-CLASS ID* (**1**) | *RA* (**2**) | *DEC* (**2**) | $z_\lambda$ | $z_s$ | *Member*($z_s$) (**3**) |
|---|---|---|---|---|---|
| 40 | 35.189 | -3.434 | 0.35 | 0.33 | 11 |
| 54 | 145.938 | 16.738 | 0.18 | 0.17 | 3 |
| 56 | 145.886 | 16.667 | 0.28 | 0.26 | 3 |
| 62 | 44.142 | 0.103 | 0.37 | 0.36 | 16 |
| 78 | 10.722 | -9.570 | 0.42 | 0.42 | 1 |
| 88 | 183.395 | 2.896 | 0.41 | 0.41 | 2 |
| 96 | 9.276 | 9.158 | 0.27 | 0.25 | 7 |
| 99 | 28.176 | 1.009 | 0.23 | 0.22 | 21 |
| 109 | 10.961 | 0.792 | 0.47 | 0.48 | 8 |
| 110 | 10.720 | 0.714 | 0.27 | 0.28 | 15 |
| 156 | 187.580 | 16.281 | 0.20 | 0.20 | 3 |
| 169 | 35.522 | -4.549 | 0.32 | 0.32 | 8 |
| 201 | 328.406 | 17.696 | 0.25 | 0.23 | 5 |
| 229 | 148.582 | 17.597 | 0.37 | 0.38 | 3 |
| 264 | 213.602 | -0.379 | 0.14 | 0.14 | 18 |
| 336 | 150.772 | 32.897 | 0.41 | 0.42 | 4 |
| 342 | 36.455 | -5.897 | 0.22 | 0.20 | 4 |
| 343 | 33.872 | -4.681 | 0.35 | 0.35 | 14 |
| 347 | 35.486 | -5.757 | 0.26 | 0.26 | 2 |
| 377 | 6.648 | 17.159 | 0.40 | 0.40 | 7 |
| 382 | 180.204 | -3.458 | 0.37 | 0.40 | 1 |
| 402 | 223.232 | 16.702 | 0.06 | 0.05 | 73 |
| 403 | 233.135 | 4.677 | 0.06 | 0.04 | 34 |
| 458 | 4.638 | 16.436 | 0.57 | 0.55 | 2 |
| 466 | 202.704 | -1.865 | 0.10 | 0.09 | 33 |
| 468 | 202.772 | -1.765 | 0.55 | 0.56 | 5 |
| 470 | 208.572 | -2.366 | 0.57 | 0.55 | 4 |
| 561 | 229.078 | 0.092 | 0.12 | 0.12 | 22 |
| 562 | 229.102 | -0.832 | 0.38 | 0.38 | 2 |
| 564 | 229.185 | -0.973 | 0.12 | 0.12 | 25 |
| 574 | 7.640 | 26.303 | 0.50 | 0.50 | 6 |
| 632 | 190.391 | 32.841 | 0.36 | 0.40 | 6 |
| 653 | 173.314 | 66.376 | 0.12 | 0.11 | 33 |
| 686 | 131.762 | 34.854 | 0.48 | 0.46 | 1 |
| 706 | 170.030 | 43.301 | 0.58 | 0.61 | 3 |
| 740 | 182.814 | 39.195 | 0.34 | 0.35 | 11 |
| 787 | 196.001 | 67.515 | 0.22 | 0.23 | 16 |
| 841 | 140.259 | 30.092 | 0.55 | 0.55 | 4 |
| 870 | 230.776 | 8.609 | 0.06 | 0.04 | 65 |
| 890 | 20.273 | 3.802 | 0.34 | 0.36 | 2 |
| 1059 | 358.902 | 5.855 | 0.28 | 0.28 | 18 |
| 1062 | 159.508 | 41.773 | 0.13 | 0.12 | 15 |
| 1063 | 201.287 | 65.836 | 0.18 | 0.16 | 4 |
| 1086 | 217.764 | 42.241 | 0.44 | 0.42 | 6 |
| 1159 | 134.057 | 37.936 | 0.41 | 0.41 | 13 |
| 1185 | 191.328 | 56.769 | 0.53 | 0.49 | 2 |
| 1188 | 213.894 | 28.394 | 0.24 | 0.22 | 2 |
| 1266 | 229.191 | 7.022 | 0.06 | 0.04 | 64 |
| 1282 | 197.945 | 22.028 | 0.17 | 0.17 | 7 |
| 1283 | 197.736 | 21.966 | 0.29 | 0.28 | 2 |
| 1307 | 7.370 | -0.214 | 0.06 | 0.06 | 86 |
| 1341 | 125.461 | 1.200 | 0.09 | 0.09 | 11 |
| 1368 | 154.404 | 59.563 | 0.29 | 0.28 | 19 |
| 1442 | 17.513 | 13.978 | 0.07 | 0.06 | 36 |
| 1443 | 155.544 | 38.523 | 0.07 | 0.06 | 50 |
| 1537 | 187.961 | 12.001 | 0.25 | 0.25 | 2 |
| 1544 | 121.939 | 39.771 | 0.36 | 0.37 | 4 |
| 1581 | 148.809 | 18.208 | 0.41 | 0.42 | 2 |
| 1627 | 197.135 | 53.704 | 0.35 | 0.33 | 14 |
| 1635 | 114.025 | 43.652 | 0.44 | 0.43 | 8 |
| 1637 | 154.264 | 39.049 | 0.21 | 0.21 | 22 |
| 1642 | 132.201 | 44.938 | 0.55 | 0.54 | 9 |
| 1676 | 200.182 | 33.154 | 0.29 | 0.30 | 2 |
| 1677 | 200.037 | 33.089 | 0.05 | 0.04 | 40 |
| 1678 | 145.756 | 46.993 | 0.36 | 0.40 | 13 |
| 1764 | 337.052 | 20.583 | 0.38 | 0.41 | 6 |
| 1813 | 164.233 | 6.978 | 0.32 | 0.30 | 1 |
| 1853 | 350.358 | 19.753 | 0.33 | 0.30 | 24 |
| 1862 | 190.793 | 14.340 | 0.34 | 0.34 | 1 |
| 1931 | 196.957 | 29.429 | 0.26 | 0.24 | 2 |
| 1944 | 149.044 | -0.365 | 0.59 | 0.59 | 3 |
| 2022 | 215.001 | 6.581 | 0.56 | 0.56 | 3 |
| 2051 | 179.602 | 44.091 | 0.40 | 0.41 | 1 |
| 2080 | 139.896 | 30.531 | 0.42 | 0.43 | 3 |
| 2081 | 140.220 | 30.466 | 0.29 | 0.29 | 5 |
| 2090 | 335.987 | -1.583 | 0.10 | 0.09 | 7 |
| 2093 | 335.812 | -1.661 | 0.30 | 0.30 | 1 |
| 2109 | 4.406 | -0.877 | 0.21 | 0.20 | 25 |
| 2113 | 188.483 | 15.437 | 0.23 | 0.23 | 2 |
| 2116 | 188.570 | 15.252 | 0.28 | 0.29 | 1 |
| 2129 | 329.371 | -7.800 | 0.07 | 0.06 | 58 |
| 2155 | 152.835 | 53.572 | 0.41 | 0.39 | 6 |
| 2202 | 340.841 | -9.597 | 0.44 | 0.44 | 2 |
| 2209 | 149.769 | 13.089 | 0.39 | 0.40 | 1 |
| 2211 | 189.670 | 9.475 | 0.23 | 0.23 | 3 |
| 2214 | 126.054 | 30.077 | 0.32 | 0.30 | 1 |
| 2295 | 120.607 | 39.091 | 0.35 | 0.37 | 2 |
| 2317 | 227.368 | 7.557 | 0.08 | 0.07 | 41 |
| 419 | 337.096 | -5.342 | 0.35 | - | - |
| 535 | 339.914 | -5.725 | 0.26 | - | - |
| 2020 | 214.847 | 6.643 | 0.55 | - | - |
| 2130 | 329.308 | -7.712 | 0.48 | - | - |

(**1**) The X-ray and optical images are retrievable from the public X-CLASS database http://xmm-lss.in2p3.fr:8080/l4sdb/
(**2**) RA and DEC are J2000 coordinates in degrees.
(**3**) The number of cluster members with spectroscopic redshifts.

Table A2: Characteristic X-ray properties of 57 galaxy cluster measured within $R_{500}$.

| *X-CLASS ID* | $T$ keV | $-eT$ | $+eT$ | $R_{500}$ Mpc | $-eR_{500}$ | $+eR_{500}$ | $F_x$ [0.5-2 keV] $10^{-14}$ (erg s$^{-1}$cm$^{-2}$) | $-eF_x$ | $+eF_x$ | $L_x^{\star}$ [0.5-2 keV] erg s$^{-1}$ | $-eL_x$ | $+eL_x$ |
|---|---|---|---|---|---|---|---|---|---|---|---|---|
| 40 | 7.24 | 2.31 | 5.24 | 1.14 | 0.22 | 0.41 | 8.52 | 0.80 | 0.41 | 43.45 | 0.032 | 0.031 |
| 56 | 5.52 | 0.84 | 0.94 | 0.99 | 0.08 | 0.09 | 23.18 | 0.87 | 0.77 | 43.65 | 0.031 | 0.016 |
| 62 | 7.35 | 0.46 | 0.46 | 1.14 | 0.04 | 0.04 | 61.41 | 0.73 | 0.74 | 44.42 | 0.006 | 0.006 |
| 88 | 4.39 | 0.48 | 0.42 | 0.83 | 0.03 | 0.06 | 18.64 | 0.56 | 0.36 | 44.01 | 0.014 | 0.011 |
| 110 | 2.16 | 0.38 | 0.48 | 0.64 | 0.06 | 0.07 | 4.51 | 0.22 | 0.43 | 43.05 | 0.040 | 0.036 |
| 156 | 5.26 | 0.57 | 0.87 | 1.09 | 0.07 | 0.10 | 43.89 | 1.24 | 1.30 | 43.70 | 0.014 | 0.013 |
| 169 | 2.64 | 1.17 | 2.64 | 0.51 | 0.26 | 0.37 | 2.48 | 0.23 | 0.19 | 42.90 | 0.049 | 0.042 |
| 229 | 2.92 | 0.26 | 0.25 | 0.70 | 0.04 | 0.03 | 24.91 | 0.67 | 0.59 | 44.09 | 0.012 | 0.012 |
| 264 | 2.62 | 0.45 | 0.27 | 0.78 | 0.07 | 0.04 | 50.13 | 1.29 | 1.25 | 43.41 | 0.011 | 0.011 |
| 336 | 4.05 | 0.17 | 0.17 | 0.82 | 0.02 | 0.02 | 28.49 | 0.39 | 0.44 | 44.22 | 0.007 | 0.007 |
| 342 | 4.14 | 1.43 | 1.70 | 1.09 | 0.20 | 0.33 | 25.58 | 1.34 | 0.94 | 43.47 | 0.022 | 0.029 |
| 347 | 1.82 | 0.38 | 0.54 | 0.54 | 0.04 | 0.12 | 6.10 | 0.43 | 0.79 | 43.11 | 0.042 | 0.040 |
| 377 | 3.91 | 0.10 | 0.11 | 0.80 | 0.01 | 0.01 | 26.47 | 0.25 | 0.23 | 44.15 | 0.005 | 0.005 |
| 382 | 5.96 | 0.38 | 0.38 | 1.09 | 0.04 | 0.04 | 22.33 | 0.32 | 0.42 | 44.04 | 0.008 | 0.008 |
| 458 | 10.13 | 0.25 | 0.25 | 1.22 | 0.02 | 0.02 | 111.34 | 0.64 | 0.66 | 45.05 | 0.003 | 0.003 |
| 468 | 6.73 | 0.39 | 0.73 | 1.01 | 0.05 | 0.04 | 18.08 | 0.29 | 0.28 | 44.28 | 0.012 | 0.010 |
| 470 | 6.19 | 0.75 | 0.93 | 0.92 | 0.06 | 0.07 | 13.53 | 0.42 | 0.26 | 44.15 | 0.014 | 0.013 |
| 562 | 4.74 | 0.80 | 1.05 | 0.88 | 0.09 | 0.09 | 9.49 | 0.56 | 0.38 | 43.65 | 0.026 | 0.025 |
| 574 | 6.64 | 0.42 | 0.56 | 1.01 | 0.04 | 0.05 | 19.41 | 0.28 | 0.37 | 44.21 | 0.005 | 0.009 |
| 632 | 6.50 | 0.27 | 0.28 | 1.14 | 0.03 | 0.03 | 53.72 | 0.54 | 0.60 | 44.42 | 0.006 | 0.006 |
| 706 | 5.62 | 0.31 | 0.29 | 0.87 | 0.02 | 0.03 | 29.76 | 0.40 | 0.46 | 44.59 | 0.007 | 0.007 |
| 740 | 5.57 | 0.49 | 0.44 | 1.02 | 0.03 | 0.06 | 34.21 | 0.66 | 0.53 | 44.12 | 0.009 | 0.011 |
| 787 | 3.36 | 0.11 | 0.19 | 0.84 | 0.02 | 0.02 | 54.10 | 0.52 | 0.60 | 43.92 | 0.006 | 0.005 |
| 841 | 2.01 | 0.36 | 0.64 | 0.51 | 0.05 | 0.08 | 2.94 | 0.28 | 0.27 | 43.56 | 0.042 | 0.040 |
| 890 | 4.97 | 0.91 | 1.79 | 0.97 | 0.11 | 0.16 | 15.39 | 0.66 | 0.41 | 43.79 | 0.017 | 0.016 |
| 1063 | 1.28 | 0.04 | 0.03 | 0.50 | 0.01 | 0.01 | 8.26 | 0.24 | 0.27 | 42.80 | 0.015 | 0.015 |
| 1086 | 4.27 | 2.28 | 3.60 | 0.93 | 0.24 | 0.49 | 4.58 | 0.94 | 0.23 | 43.42 | 0.059 | 0.081 |
| 1159 | 7.43 | 0.36 | 0.37 | 1.14 | 0.03 | 0.03 | 55.18 | 0.54 | 0.47 | 44.47 | 0.005 | 0.005 |
| 1188 | 1.75 | 0.12 | 0.40 | 0.56 | 0.02 | 0.07 | 7.23 | 0.48 | 0.83 | 43.04 | 0.040 | 0.037 |
| 1283 | 6.76 | 0.93 | 1.64 | 1.18 | 0.12 | 0.12 | 36.65 | 1.17 | 0.57 | 43.92 | 0.017 | 0.012 |
| 1544 | 8.37 | 1.48 | 2.72 | 1.24 | 0.12 | 0.19 | 30.50 | 0.96 | 1.32 | 44.11 | 0.018 | 0.020 |
| 1627 | 3.09 | 0.18 | 0.18 | 0.70 | 0.02 | 0.02 | 13.26 | 0.22 | 0.22 | 43.66 | 0.010 | 0.010 |
| 1635 | 4.60 | 0.56 | 0.66 | 0.89 | 0.06 | 0.07 | 12.94 | 0.45 | 0.33 | 43.93 | 0.017 | 0.017 |
| 1642 | 2.12 | 0.35 | 0.34 | 0.56 | 0.05 | 0.04 | 3.84 | 0.17 | 0.18 | 43.66 | 0.024 | 0.026 |
| 1676 | 2.53 | 0.18 | 0.18 | 0.68 | 0.03 | 0.03 | 10.13 | 0.30 | 0.26 | 43.45 | 0.013 | 0.013 |
| 1764 | 5.26 | 0.67 | 1.13 | 1.01 | 0.07 | 0.15 | 8.37 | 0.32 | 0.27 | 43.68 | 0.021 | 0.018 |
| 1853 | 4.58 | 0.36 | 0.39 | 0.88 | 0.04 | 0.04 | 21.56 | 0.46 | 0.42 | 43.80 | 0.010 | 0.010 |
| 1862 | 2.25 | 0.28 | 0.47 | 0.56 | 0.05 | 0.04 | 4.10 | 0.23 | 0.25 | 43.21 | 0.027 | 0.026 |
| 1931 | 3.01 | 0.14 | 0.14 | 0.73 | 0.02 | 0.02 | 36.20 | 0.48 | 0.46 | 43.79 | 0.006 | 0.006 |
| 1944 | 3.88 | 1.03 | 1.37 | 0.71 | 0.11 | 0.12 | 3.52 | 0.24 | 0.19 | 43.66 | 0.039 | 0.039 |
| 2022 | 3.67 | 0.28 | 0.26 | 0.71 | 0.03 | 0.03 | 7.88 | 0.20 | 0.16 | 43.96 | 0.011 | 0.011 |
| 2051 | 4.24 | 0.83 | 1.18 | 0.84 | 0.09 | 0.12 | 4.45 | 0.32 | 0.23 | 43.38 | 0.028 | 0.026 |
| 2080 | 3.39 | 0.25 | 0.19 | 0.75 | 0.03 | 0.02 | 18.56 | 0.38 | 0.22 | 44.06 | 0.007 | 0.009 |
| 2081 | 3.20 | 0.14 | 0.14 | 0.77 | 0.02 | 0.02 | 31.02 | 0.49 | 0.37 | 43.91 | 0.006 | 0.006 |
| 2093 | 4.14 | 0.41 | 0.52 | 0.87 | 0.05 | 0.06 | 22.97 | 0.69 | 0.66 | 43.81 | 0.015 | 0.014 |
| 2155 | 6.20 | 1.51 | 2.08 | 0.98 | 0.13 | 0.16 | 7.92 | 0.44 | 0.26 | 43.56 | 0.025 | 0.023 |
| 2209 | 3.26 | 0.35 | 0.45 | 0.74 | 0.04 | 0.05 | 9.64 | 0.38 | 0.23 | 43.71 | 0.017 | 0.017 |
| 2214 | 2.32 | 0.26 | 0.78 | 0.56 | 0.04 | 0.08 | 4.88 | 0.31 | 0.21 | 43.17 | 0.025 | 0.028 |
| 2295 | 2.77 | 0.23 | 0.46 | 0.71 | 0.03 | 0.06 | 24.55 | 1.06 | 1.00 | 44.07 | 0.020 | 0.019 |
| 1678 | 5.73 | 0.15 | 0.15 | 1.07 | 0.02 | 0.02 | 59.17 | 0.41 | 0.43 | 44.47 | 0.003 | 0.003 |
| 2109 | 4.70 | 0.18 | 0.19 | 0.96 | 0.02 | 0.02 | 73.73 | 0.67 | 0.68 | 43.94 | 0.004 | 0.004 |
| 2202 | 8.12 | 0.08 | 0.08 | 1.18 | 0.01 | 0.01 | 153.06 | 0.36 | 0.44 | 44.99 | 0.001 | 0.001 |
| 2211 | 3.72 | 0.19 | 0.19 | 0.86 | 0.02 | 0.02 | 40.80 | 0.58 | 0.54 | 43.78 | 0.003 | 0.006 |
| 419 | 4.20 | 0.87 | 0.66 | 0.92 | 0.10 | 0.07 | 7.00 | 0.25 | 0.25 | 43.45 | 0.018 | 0.023 |
| 535 | 3.00 | 0.25 | 0.25 | 0.74 | 0.05 | 0.03 | 24.95 | 0.71 | 0.49 | 43.71 | 0.012 | 0.011 |
| 2020 | 4.82 | 0.43 | 0.44 | 0.83 | 0.04 | 0.04 | 12.67 | 0.32 | 0.21 | 44.13 | 0.013 | 0.013 |
| 2130 | 1.87 | 0.19 | 0.25 | 0.51 | 0.03 | 0.04 | 4.17 | 0.22 | 0.34 | 43.58 | 0.030 | 0.028 |

Table A3: Characteristic X-ray properties of 70 galaxy cluster measured within $R_{fit}$.

| *X-CLASS ID* | $R_{fit}$ arcsec | $T$ keV | $-eT$ | $+eT$ | $F_x$ [0.5-2 keV] $10^{-14}$ erg s$^{-1}$cm$^{-2}$ | $-eF_x$ | $+eF_x$ | $L_x^{\star}$ [0.5-2 keV] erg s$^{-1}$ | $-eL_x$ | $+eL_x$ |
|---|---|---|---|---|---|---|---|---|---|---|
| 40 | 75.0 | 2.07 | 0.25 | 0.41 | 3.79 | 0.26 | 0.24 | 43.15 | 0.028 | 0.027 |
| 54 | 150.0 | 2.55 | 0.24 | 0.35 | 10.77 | 0.42 | 0.27 | 42.92 | 0.016 | 0.016 |
| 56 | 212.5 | 5.07 | 0.61 | 0.95 | 21.66 | 0.79 | 0.69 | 43.62 | 0.015 | 0.015 |
| 62 | 212.5 | 6.86 | 0.36 | 0.49 | 60.10 | 0.94 | 0.57 | 44.41 | 0.006 | 0.007 |
| 78 | 75.0 | 8.68 | 2.53 | 6.69 | 4.29 | 0.54 | 0.22 | 43.38 | 0.040 | 0.040 |
| 88 | 162.5 | 4.17 | 0.32 | 0.56 | 18.96 | 0.34 | 0.45 | 44.01 | 0.013 | 0.012 |
| 109 | 87.5 | 3.79 | 0.74 | 1.16 | 3.90 | 0.26 | 0.24 | 43.49 | 0.033 | 0.031 |
| 110 | 137.5 | 3.05 | 0.73 | 1.18 | 4.73 | 0.45 | 0.29 | 43.05 | 0.036 | 0.034 |
| 156 | 275.5 | 5.10 | 0.56 | 0.70 | 41.75 | 1.38 | 1.11 | 43.67 | 0.013 | 0.013 |
| 169 | 75.0 | 5.32 | 1.54 | 3.03 | 2.38 | 0.19 | 0.08 | 42.87 | 0.040 | 0.036 |
| 229 | 187.5 | 2.94 | 0.25 | 0.25 | 29.70 | 0.52 | 0.77 | 44.17 | 0.012 | 0.012 |
| 264 | 237.5 | 2.94 | 0.27 | 0.23 | 42.13 | 0.92 | 0.89 | 43.33 | 0.010 | 0.010 |
| 336 | 212.5 | 4.59 | 0.25 | 0.25 | 33.37 | 0.38 | 0.51 | 44.28 | 0.007 | 0.007 |
| 342 | 150.0 | 3.11 | 0.48 | 0.56 | 14.89 | 0.76 | 0.63 | 43.25 | 0.021 | 0.020 |
| 343 | 125.0 | 4.67 | 0.74 | 1.08 | 12.03 | 0.67 | 0.53 | 43.66 | 0.021 | 0.021 |
| 347 | 112.5 | 3.79 | 1.43 | 2.04 | 6.05 | 0.49 | 0.55 | 43.08 | 0.040 | 0.035 |
| 377 | 225.0 | 3.66 | 0.11 | 0.11 | 32.67 | 0.29 | 0.28 | 44.24 | 0.005 | 0.005 |
| 382 | 175.0 | 6.25 | 0.37 | 0.37 | 20.42 | 0.31 | 0.31 | 44.00 | 0.008 | 0.008 |
| 468 | 62.5 | 5.12 | 0.55 | 0.72 | 5.64 | 0.15 | 0.21 | 43.79 | 0.019 | 0.019 |
| 470 | 112.5 | 7.25 | 1.09 | 1.08 | 11.28 | 0.39 | 0.16 | 44.07 | 0.014 | 0.013 |
| 562 | 112.5 | 3.39 | 0.31 | 0.44 | 9.12 | 0.40 | 0.26 | 43.65 | 0.018 | 0.017 |
| 574 | 175.0 | 6.69 | 0.42 | 0.60 | 20.24 | 0.28 | 0.36 | 44.23 | 0.009 | 0.009 |
| 632 | 225.0 | 6.30 | 0.27 | 0.28 | 55.16 | 0.61 | 0.64 | 44.43 | 0.006 | 0.006 |
| 686 | 62.5 | 3.91 | 1.50 | 2.65 | 3.47 | 0.54 | 0.29 | 43.41 | 0.055 | 0.054 |
| 706 | 225.0 | 6.06 | 0.31 | 0.32 | 37.94 | 0.58 | 0.49 | 44.69 | 0.007 | 0.007 |
| 740 | 162.5 | 5.80 | 0.43 | 0.44 | 30.13 | 0.54 | 0.46 | 44.06 | 0.009 | 0.009 |
| 787 | 200.0 | 3.62 | 0.15 | 0.15 | 48.15 | 0.58 | 0.61 | 43.86 | 0.006 | 0.006 |
| 841 | 78.5 | 1.99 | 0.33 | 0.56 | 2.96 | 0.25 | 0.31 | 43.57 | 0.042 | 0.040 |
| 890 | 150.0 | 4.87 | 0.87 | 1.31 | 12.46 | 0.43 | 0.33 | 43.70 | 0.016 | 0.016 |
| 1059 | 112.5 | 13.56 | 4.60 | 8.49 | 5.54 | 0.41 | 0.27 | 43.10 | 0.028 | 0.027 |
| 1062 | 225.0 | 3.11 | 0.22 | 0.22 | 25.36 | 0.38 | 0.67 | 43.00 | 0.009 | 0.009 |
| 1063 | 112.5 | 1.28 | 0.04 | 0.04 | 5.99 | 0.20 | 0.16 | 42.66 | 0.016 | 0.015 |
| 1086 | 87.5 | 5.31 | 1.24 | 1.33 | 5.92 | 0.47 | 0.40 | 43.53 | 0.030 | 0.031 |
| 1159 | 225.0 | 7.58 | 0.37 | 0.37 | 57.11 | 0.71 | 0.47 | 44.49 | 0.005 | 0.005 |
| 1185 | 87.5 | 27.88 | 18.95 | -27.30 | 3.08 | 3.08 | 0.34 | 43.34 | 0.065 | 0.062 |
| 1188 | 87.5 | 3.13 | 0.59 | 0.64 | 8.16 | 0.42 | 0.38 | 43.07 | 0.024 | 0.023 |
| 1282 | 237.5 | 3.87 | 0.19 | 0.19 | 52.13 | 0.79 | 0.69 | 43.61 | 0.007 | 0.006 |
| 1283 | 225.0 | 6.63 | 0.82 | 1.19 | 34.97 | 1.11 | 0.62 | 43.90 | 0.014 | 0.013 |
| 1307 | 162.5 | 5.07 | 0.62 | 1.13 | 34.44 | 0.88 | 0.73 | 42.48 | 0.011 | 0.011 |
| 1537 | 62.5 | 4.58 | 0.44 | 0.53 | 15.59 | 0.42 | 0.49 | 43.45 | 0.016 | 0.015 |
| 1544 | 225.0 | 10.38 | 2.05 | 4.58 | 29.95 | 1.04 | 0.79 | 44.10 | 0.018 | 0.021 |
| 1581 | 37.5 | 7.57 | 2.96 | 8.19 | 1.87 | 0.25 | 0.11 | 43.01 | 0.049 | 0.046 |
| 1627 | 162.5 | 3.15 | 0.18 | 0.19 | 14.22 | 0.23 | 0.23 | 43.69 | 0.010 | 0.010 |
| 1635 | 150.0 | 4.66 | 0.55 | 0.64 | 12.93 | 0.37 | 0.36 | 43.93 | 0.017 | 0.016 |
| 1642 | 112.5 | 2.55 | 0.35 | 0.73 | 4.72 | 0.30 | 0.22 | 43.74 | 0.030 | 0.024 |
| 1676 | 112.5 | 2.96 | 0.26 | 0.26 | 8.13 | 0.23 | 0.20 | 43.35 | 0.014 | 0.014 |
| 1764 | 162.5 | 4.57 | 0.87 | 0.70 | 7.47 | 0.25 | 0.21 | 43.63 | 0.019 | 0.024 |
| 1813 | 137.5 | 11.77 | 4.19 | 33.04 | 4.74 | 0.63 | 0.38 | 43.08 | 0.044 | 0.031 |
| 1853 | 187.5 | 4.48 | 0.33 | 0.38 | 22.01 | 0.37 | 0.52 | 43.80 | 0.010 | 0.010 |
| 1862 | 75.0 | 1.58 | 0.10 | 0.10 | 3.54 | 0.18 | 0.22 | 43.16 | 0.023 | 0.023 |
| 1931 | 287.5 | 2.89 | 0.17 | 0.17 | 38.10 | 0.52 | 0.49 | 43.82 | 0.007 | 0.007 |
| 1944 | 100.0 | 3.26 | 0.63 | 1.22 | 3.47 | 0.37 | 0.17 | 43.66 | 0.035 | 0.040 |
| 2022 | 125.0 | 3.34 | 0.19 | 0.42 | 8.49 | 0.21 | 0.19 | 44.00 | 0.015 | 0.011 |
| 2051 | 100.0 | 3.16 | 0.40 | 0.54 | 4.55 | 0.22 | 0.17 | 43.40 | 0.019 | 0.019 |
| 2080 | 125.0 | 3.75 | 0.18 | 0.18 | 18.91 | 0.27 | 0.19 | 44.06 | 0.007 | 0.007 |
| 2081 | 300.0 | 2.87 | 0.17 | 0.18 | 35.37 | 0.64 | 0.54 | 43.97 | 0.004 | 0.007 |
| 2093 | 150.0 | 4.04 | 0.32 | 0.35 | 22.96 | 0.66 | 0.59 | 43.81 | 0.012 | 0.011 |
| 2116 | 275.0 | 4.09 | 0.34 | 0.45 | 28.55 | 0.76 | 0.72 | 43.85 | 0.012 | 0.011 |
| 2155 | 125.0 | 9.78 | 2.73 | 4.38 | 6.46 | 0.43 | 0.22 | 43.46 | 0.022 | 0.022 |
| 2209 | 162.5 | 3.60 | 0.50 | 0.54 | 10.19 | 0.41 | 0.29 | 43.73 | 0.018 | 0.019 |
| 2214 | 150.0 | 1.57 | 0.14 | 0.13 | 5.46 | 0.24 | 0.40 | 43.24 | 0.028 | 0.027 |

Table A3: Characteristic X-ray properties of 70 galaxy cluster measured within $R_{fit}$.

| *X-CLASS ID* | $R_{fit}$ arcsec | $T$ keV | $-eT$ | $+eT$ | $F_x$ [0.5-2 keV] $10^{-14}$ erg s$^{-1}$cm$^{-2}$ | $-eF_x$ | $+eF_x$ | $L_x^\star$ [0.5-2 keV] erg s$^{-1}$ | $-eL_x$ | $+eL_x$ |
|---|---|---|---|---|---|---|---|---|---|---|
| 2295 | 200.0 | 2.16 | 0.15 | 0.18 | 37.51 | 1.48 | 1.96 | 44.27 | 0.017 | 0.019 |
| 1678 | 200.0 | 5.72 | 0.15 | 0.15 | 59.36 | 0.40 | 0.41 | 44.47 | 0.003 | 0.003 |
| 2109 | 225.0 | 4.53 | 0.17 | 0.17 | 63.71 | 0.57 | 0.45 | 43.88 | 0.004 | 0.004 |
| 2113 | 250.0 | 5.13 | 0.13 | 0.13 | 84.79 | 0.54 | 0.57 | 44.12 | 0.003 | 0.003 |
| 1677 | 75.0 | 0.96 | 0.02 | 0.02 | 4.75 | 0.16 | 0.12 | 41.22 | 0.014 | 0.014 |
| 419 | 112.5 | 3.38 | 0.35 | 0.54 | 4.79 | 0.17 | 0.12 | 43.30 | 0.021 | 0.019 |
| 535 | 300.0 | 2.03 | 0.11 | 0.11 | 34.51 | 1.20 | 0.75 | 43.87 | 0.012 | 0.011 |
| 2020 | 137.5 | 4.55 | 0.45 | 0.47 | 12.58 | 0.33 | 0.30 | 44.13 | 0.014 | 0.014 |
| 2130 | 62.5 | 2.00 | 0.21 | 0.24 | 3.65 | 0.24 | 0.18 | 43.52 | 0.027 | 0.026 |

Table A4: Characteristic X-ray properties of 92 galaxy cluster measured within $R_{300kpc}$.

| *X-CLASS ID* | $R_{300kpc}$ arcsec | $T$ keV | $-eT$ | $+eT$ | $F_x$ [0.5-2 keV] $10^{-14}$ erg s$^{-1}$cm$^{-2}$ | $-eF_x$ | $+eF_x$ | $L_x^\star$ [0.5-2 keV] erg s$^{-1}$ | $-eL_x$ | $+eL_x$ |
|---|---|---|---|---|---|---|---|---|---|---|
| 40 | 60.34 | 1.93 | 0.22 | 0.27 | 3.27 | 0.18 | 0.22 | 43.09 | 0.029 | 0.028 |
| 54 | 96.87 | 2.22 | 0.15 | 0.35 | 7.69 | 0.27 | 0.25 | 42.78 | 0.017 | 0.016 |
| 56 | 71.18 | 3.02 | 0.33 | 0.37 | 7.25 | 0.34 | 0.29 | 43.16 | 0.019 | 0.019 |
| 62 | 58.48 | 5.64 | 0.30 | 0.34 | 29.38 | 0.46 | 0.36 | 44.11 | 0.008 | 0.008 |
| 78 | 54.11 | 8.26 | 2.74 | 6.84 | 3.24 | 0.35 | 0.22 | 43.26 | 0.022 | 0.043 |
| 88 | 54.90 | 5.15 | 0.49 | 0.56 | 7.48 | 0.24 | 0.20 | 43.60 | 0.015 | 0.015 |
| 96 | 72.69 | 10.4 | 0.31 | 0.31 | 97.31 | 0.58 | 0.69 | 44.26 | 0.003 | 0.003 |
| 99 | 82.29 | 6.32 | 0.15 | 0.15 | 134.3 | 0.80 | 1.00 | 44.26 | 0.003 | 0.003 |
| 109 | 50.80 | 3.42 | 0.70 | 1.06 | 2.28 | 0.18 | 0.15 | 43.26 | 0.039 | 0.038 |
| 110 | 72.24 | 3.57 | 0.81 | 1.24 | 2.85 | 0.26 | 0.18 | 42.83 | 0.035 | 0.035 |
| 156 | 91.94 | 5.32 | 0.51 | 0.84 | 18.22 | 0.73 | 0.52 | 43.31 | 0.015 | 0.014 |
| 169 | 64.12 | 10.47 | 4.28 | 9.03 | 2.06 | 0.19 | 0.17 | 42.78 | 0.040 | 0.021 |
| 201 | 76.73 | 11.35 | 0.40 | 0.40 | 326.59 | 1.29 | 1.71 | 44.71 | 0.002 | 0.002 |
| 229 | 58.32 | 2.23 | 0.28 | 0.61 | 3.10 | 0.23 | 0.17 | 43.21 | 0.034 | 0.032 |
| 264 | 123.26 | 2.61 | 0.23 | 0.44 | 13.67 | 0.34 | 0.51 | 42.85 | 0.015 | 0.015 |
| 336 | 55.02 | 4.43 | 0.21 | 0.22 | 18.47 | 0.34 | 0.18 | 44.02 | 0.008 | 0.008 |
| 342 | 83.60 | 3.19 | 0.41 | 0.57 | 10.32 | 0.47 | 0.28 | 43.09 | 0.020 | 0.073 |
| 343 | 60.49 | 3.57 | 0.55 | 0.61 | 7.27 | 0.32 | 0.25 | 43.45 | 0.022 | 0.021 |
| 347 | 74.23 | 2.17 | 0.35 | 0.61 | 4.58 | 0.29 | 0.43 | 42.98 | 0.038 | 0.036 |
| 377 | 55.69 | 4.24 | 0.12 | 0.15 | 15.17 | 0.19 | 0.14 | 43.90 | 0.006 | 0.006 |
| 382 | 58.96 | 5.28 | 0.31 | 0.45 | 9.44 | 0.19 | 0.15 | 43.67 | 0.010 | 0.009 |
| 402 | 271.03 | 2.05 | 0.02 | 0.02 | 235.23 | 0.73 | 0.97 | 43.09 | 0.002 | 0.002 |
| 403 | 242.45 | 2.11 | 0.02 | 0.02 | 158.43 | 1.03 | 0.57 | 42.80 | 0.002 | 0.002 |
| 458 | 45.93 | 10.80 | 0.36 | 0.53 | 41.23 | 0.33 | 0.36 | 44.62 | 0.004 | 0.004 |
| 466 | 170.13 | 4.36 | 0.07 | 0.07 | 186.16 | 1.06 | 0.64 | 43.56 | 0.002 | 0.002 |
| 468 | 46.69 | 4.45 | 0.53 | 0.62 | 5.04 | 0.20 | 0.18 | 43.75 | 0.021 | 0.021 |
| 470 | 45.91 | 8.73 | 0.88 | 1.71 | 5.06 | 0.15 | 0.14 | 43.71 | 0.017 | 0.015 |
| 561 | 137.83 | 5.80 | 0.10 | 0.10 | 203.54 | 0.94 | 0.76 | 43.90 | 0.002 | 0.002 |
| 562 | 58.10 | 2.60 | 0.18 | 0.21 | 7.15 | 0.31 | 0.17 | 43.56 | 0.015 | 0.015 |
| 564 | 141.58 | 4.87 | 0.12 | 0.12 | 82.88 | 0.63 | 0.74 | 43.51 | 0.003 | 0.003 |
| 574 | 49.05 | 6.21 | 0.43 | 0.44 | 9.12 | 0.16 | 0.17 | 43.89 | 0.010 | 0.010 |
| 632 | 59.37 | 6.28 | 0.38 | 0.39 | 18.17 | 0.31 | 0.3 | 43.95 | 0.009 | 0.008 |
| 653 | 137.64 | 5.15 | 0.12 | 0.12 | 206.91 | 1.21 | 1.19 | 43.83 | 0.003 | 0.003 |
| 686 | 50.33 | 2.05 | 0.39 | 0.73 | 3.02 | 0.23 | 0.25 | 43.40 | 0.046 | 0.043 |
| 706 | 45.43 | 5.58 | 0.26 | 0.31 | 17.46 | 0.29 | 0.27 | 44.36 | 0.004 | 0.008 |
| 740 | 61.90 | 5.64 | 0.45 | 0.47 | 17.54 | 0.32 | 0.37 | 43.83 | 0.011 | 0.011 |
| 787 | 84.25 | 6.67 | 0.41 | 0.54 | 20.40 | 0.40 | 0.37 | 43.47 | 0.008 | 0.008 |
| 841 | 46.83 | 2.51 | 0.47 | 0.73 | 2.15 | 0.18 | 0.19 | 43.41 | 0.046 | 0.043 |
| 890 | 61.76 | 4.37 | 0.44 | 0.69 | 7.15 | 0.24 | 0.23 | 43.46 | 0.016 | 0.015 |
| 1059 | 70.40 | 3.26 | 0.69 | 1.26 | 3.21 | 0.24 | 0.21 | 42.91 | 0.036 | 0.031 |
| 1062 | 127.43 | 2.50 | 0.17 | 0.17 | 15.40 | 0.46 | 0.21 | 42.79 | 0.011 | 0.011 |
| 1063 | 98.67 | 1.22 | 0.05 | 0.04 | 4.16 | 0.18 | 0.16 | 42.50 | 0.018 | 0.017 |

Table A4: Characteristic X-ray properties of 92 galaxy cluster measured within $R_{300kpc}$.

| *X-CLASS ID* | $R_{300kpc}$ arcsec | $T$ keV | -$eT$ | +$eT$ | $F_x$ [0.5-2 keV] $10^{-14}$ erg s$^{-1}$cm$^{-2}$ | -$eF_x$ | +$eF_x$ | $L_x$ ⋆ [0.5-2 keV] erg s$^{-1}$ | -$eL_x$ | +$eL_x$ |
|---|---|---|---|---|---|---|---|---|---|---|
| 1086 | 52.61 | 2.96 | 0.71 | 0.88 | 4.44 | 0.43 | 0.27 | 43.43 | 0.035 | 0.036 |
| 1159 | 55.07 | 8.21 | 0.48 | 0.57 | 18.90 | 0.33 | 0.32 | 44.00 | 0.008 | 0.008 |
| 1185 | 47.80 | 22.64 | 14.93 | -22.64 | 1.91 | 1.91 | 0.29 | 43.13 | 0.079 | 0.069 |
| 1188 | 79.91 | 2.35 | 0.40 | 0.60 | 5.17 | 0.33 | 0.32 | 42.88 | 0.028 | 0.027 |
| 1282 | 103.40 | 3.78 | 0.19 | 0.20 | 31.73 | 0.47 | 0.46 | 43.40 | 0.007 | 0.007 |
| 1283 | 69.55 | 5.23 | 0.41 | 0.63 | 26.47 | 0.71 | 0.54 | 43.78 | 0.012 | 0.010 |
| 1307 | 162.50 | 0.00 | 0.00 | 0.00 | 0.00 | 0.00 | 0.00 | 42.82 | 0.007 | 0.008 |
| 1341 | 180.97 | 3.73 | 0.10 | 0.10 | 111.20 | 1.10 | 0.80 | 43.35 | 0.004 | 0.004 |
| 1368 | 69.11 | 7.88 | 0.84 | 0.90 | 29.60 | 0.64 | 0.52 | 43.82 | 0.010 | 0.009 |
| 1442 | 226.97 | 2.92 | 0.11 | 0.11 | 87.11 | 1.23 | 0.71 | 42.93 | 0.005 | 0.005 |
| 1443 | 227.88 | 2.31 | 0.06 | 0.06 | 123.47 | 0.97 | 0.83 | 43.00 | 0.004 | 0.004 |
| 1537 | 77.30 | 4.95 | 0.59 | 0.67 | 16.60 | 0.44 | 0.70 | 43.48 | 0.017 | 0.017 |
| 1544 | 59.27 | 6.28 | 1.09 | 1.65 | 11.91 | 0.67 | 0.49 | 43.72 | 0.026 | 0.025 |
| 1581 | 55.00 | 6.26 | 2.07 | 4.44 | 2.59 | 0.35 | 0.22 | 43.15 | 0.045 | 0.041 |
| 1627 | 60.53 | 3.29 | 0.20 | 0.28 | 7.21 | 0.14 | 0.16 | 43.40 | 0.011 | 0.011 |
| 1635 | 52.85 | 4.56 | 0.42 | 0.49 | 8.07 | 0.30 | 0.25 | 43.73 | 0.015 | 0.015 |
| 1642 | 46.85 | 2.44 | 0.31 | 0.38 | 2.65 | 0.12 | 0.13 | 43.49 | 0.024 | 0.021 |
| 1676 | 68.84 | 3.18 | 0.28 | 0.31 | 5.32 | 0.19 | 0.11 | 43.16 | 0.015 | 0.015 |
| 1764 | 57.97 | 2.37 | 0.43 | 0.61 | 1.48 | 0.12 | 0.09 | 42.97 | 0.036 | 0.035 |
| 1813 | 63.92 | 3.68 | 0.74 | 1.02 | 2.86 | 0.22 | 0.17 | 42.89 | 0.031 | 0.030 |
| 1853 | 63.42 | 4.57 | 0.30 | 0.32 | 13.10 | 0.20 | 0.29 | 43.58 | 0.010 | 0.009 |
| 1862 | 62.18 | 1.54 | 0.11 | 0.10 | 3.31 | 0.20 | 0.20 | 43.13 | 0.023 | 0.023 |
| 1931 | 74.88 | 3.36 | 0.11 | 0.19 | 26.94 | 0.38 | 0.24 | 43.66 | 0.006 | 0.005 |
| 1944 | 45.31 | 3.33 | 0.61 | 0.82 | 2.15 | 0.17 | 0.13 | 43.46 | 0.035 | 0.034 |
| 2022 | 46.40 | 4.13 | 0.29 | 0.33 | 4.36 | 0.10 | 0.11 | 43.70 | 0.013 | 0.013 |
| 2051 | 55.64 | 3.69 | 0.47 | 0.44 | 3.81 | 0.14 | 0.09 | 43.32 | 0.017 | 0.016 |
| 2080 | 53.99 | 4.15 | 0.16 | 0.16 | 12.94 | 0.16 | 0.18 | 43.89 | 0.007 | 0.007 |
| 2081 | 68.92 | 3.95 | 0.20 | 0.20 | 13.83 | 0.21 | 0.22 | 43.55 | 0.008 | 0.008 |
| 2093 | 67.76 | 3.50 | 0.21 | 0.35 | 7.79 | 0.21 | 0.23 | 43.34 | 0.015 | 0.014 |
| 2116 | 70.66 | 7.94 | 1.10 | 1.40 | 11.06 | 0.36 | 0.39 | 43.42 | 0.015 | 0.015 |
| 2155 | 55.07 | 5.40 | 1.06 | 1.53 | 2.82 | 0.15 | 0.14 | 43.12 | 0.025 | 0.024 |
| 2209 | 56.70 | 0.04 | 0.43 | 0.44 | 5.58 | 0.25 | 0.21 | 43.46 | 0.018 | 0.017 |
| 2214 | 64.71 | 2.46 | 0.63 | 0.91 | 2.91 | 0.21 | 0.16 | 42.94 | 0.030 | 0.028 |
| 2295 | 60.65 | 7.82 | 1.60 | 2.12 | 12.14 | 0.67 | 0.39 | 43.71 | 0.023 | 0.022 |
| 1637 | 89.11 | 6.45 | 0.09 | 0.10 | 220.69 | 0.69 | 0.61 | 44.40 | 0.002 | 0.002 |
| 1678 | 59.80 | 5.88 | 0.24 | 0.24 | 17.88 | 0.22 | 0.20 | 43.95 | 0.006 | 0.006 |
| 2090 | 165.01 | 4.67 | 0.11 | 0.11 | 102.67 | 0.67 | 0.73 | 43.32 | 0.003 | 0.003 |
| 2109 | 86.18 | 4.25 | 0.11 | 0.14 | 42.85 | 0.43 | 0.39 | 43.71 | 0.004 | 0.005 |
| 2113 | 82.25 | 5.46 | 0.18 | 0.22 | 34.89 | 0.34 | 0.30 | 43.73 | 0.005 | 0.005 |
| 2129 | 229.67 | 3.86 | 0.11 | 0.11 | 85.59 | 0.66 | 0.70 | 42.87 | 0.004 | 0.004 |
| 2202 | 52.86 | 8.57 | 0.13 | 0.24 | 44.56 | 0.18 | 0.18 | 44.45 | 0.003 | 0.002 |
| 2211 | 82.24 | 4.09 | 0.23 | 0.24 | 15.41 | 0.40 | 0.23 | 43.36 | 0.009 | 0.009 |
| 2317 | 202.55 | 4.19 | 0.09 | 0.09 | 71.46 | 0.44 | 0.51 | 42.98 | 0.003 | 0.003 |
| 870 | 247.40 | 4.02 | 0.04 | 0.04 | 1130.10 | 5.10 | 2.90 | 43.59 | 0.001 | 0.001 |
| 1266 | 259.30 | 3.28 | 0.02 | 0.02 | 1965.10 | 4.10 | 3.90 | 43.91 | 0.001 | 0.001 |
| 1677 | 282.71 | 1.02 | 0.02 | 0.02 | 17.43 | 0.43 | 0.37 | 41.78 | 0.010 | 0.010 |
| 419 | 60.60 | 2.81 | 0.29 | 0.32 | 3.55 | 0.15 | 0.11 | 43.18 | 0.019 | 0.020 |
| 535 | 74.68 | 0.02 | 0.22 | 0.23 | 13.95 | 0.34 | 0.40 | 43.46 | 0.012 | 0.012 |
| 2020 | 46.66 | 5.57 | 0.53 | 0.50 | 7.61 | 0.14 | 0.14 | 43.91 | 0.015 | 0.012 |
| 2130 | 50.34 | 0.05 | 0.22 | 0.25 | 3.25 | 0.18 | 0.16 | 43.46 | 0.027 | 0.027 |

⋆ Luminosity and uncertainty is in a base 10 logarithmic scale

Table A5: Temperature measurements for 22 XC1 in literature.

| Published results | | | | | | Our results | | | | | |
|---|---|---|---|---|---|---|---|---|---|---|---|
| *X-CLASS ID* | $T_{published}$ keV | $-eT_{published}$ | $+eT_{published}$ | *status* | *Ref* | $T_{500}$ keV | $-eT_{500}$ | $+eT_{500}$ | $R_{500}$ arcsec | $-eR_{500}$ | $+eR_{500}$ |
| 96 | 6.50 | 0.70 | 1.00 | [0.15-1] $R_{500}$ | 6 | - | - | - | - | - | - |
| 99 | 4.60 | 0.40 | 0.50 | [0.15-1] $R_{500}$ | 6 | - | - | - | - | - | - |
| 201 | 8.68 | 0.27 | 0.29 | 90% of SI | 1 | - | - | - | - | - | - |
| 402 | 2.18 | 0.16 | 0.19 | [0.35] R200 | 2 | - | - | - | - | - | - |
| 403 | 2.58 | 0.15 | 0.15 | [0.1-0.3] R200 | 3 | - | - | - | - | - | - |
| 458 | 8.90 | 0.03 | 0.03 | no core-$R_{500}$ | 5 | 10.13 | 0.25 | 0.25 | 189.34 | 2.51 | 2.49 |
| 466 | 4.45 | 0.13 | 0.13 | 90% of SI | 1 | - | - | - | - | - | - |
| 561 | 5.25 | 0.15 | 0.16 | 90% of SI | 1,4 | - | - | - | - | - | - |
| 564 | 3.40 | 0.08 | 0.08 | [0.15-1] $R_{500}$ | 4 | - | - | - | - | - | - |
| 653 | 6.59 | 0.42 | 0.46 | 90% of SI | 1 | - | - | - | - | - | - |
| 1341 | 3.69 | 0.24 | 0.29 | 90% of SI | 1 | - | - | - | - | - | - |
| 1368 | - | - | - | No XMM Study | 0 | - | - | - | - | - | - |
| 1442 | - | - | - | No Study | 0 | - | - | - | - | - | - |
| 1443 | - | - | - | No XMM Study | 0 | - | - | - | - | - | - |
| 1637 | 6.43 | 0.19 | 0.22 | 90% of SI | 1 | - | - | - | - | - | - |
| 2090 | - | - | - | No XMM Study | 0 | - | - | - | - | - | - |
| 2129 | 2.30 | 0.06 | 0.10 | [0.15-1] $R_{500}$ | 4 | - | - | - | - | - | - |
| 2202 | 7.98 | 0.12 | 0.12 | [0.15-0.75] $R_{500}$ | 7 | 8.11 | 0.08 | 0.08 | 206.78 | 1.15 | 1.15 |
| 2211 | - | - | - | No Study | 0 | 3.72 | 0.19 | 0.19 | 236.25 | 6.63 | 6.57 |
| 2317 | - | - | - | No XMM Study | 0 | - | - | - | - | - | - |
| 870 | - | - | - | No XMM Study | 0 | - | - | - | - | - | - |
| 1266 | - | - | - | No XMM Study | 0 | - | - | - | - | - | - |

Ref: 1-Andersson et al. (2009); 2-Pratt & Arnaud (2003); 3-Pratt & Arnaud (2005); 4-Croston et al. (2008); 5-Kotov & Vikhlinin (2005); 6-Maughan et al. (2008); 7-Planck Collaboration et al. (2011)

## APPENDIX B: R500 REPRESENTATIVE PLOTS

We present here an example of the representative plot we created through the $R_{500}$ calculation pipeline.

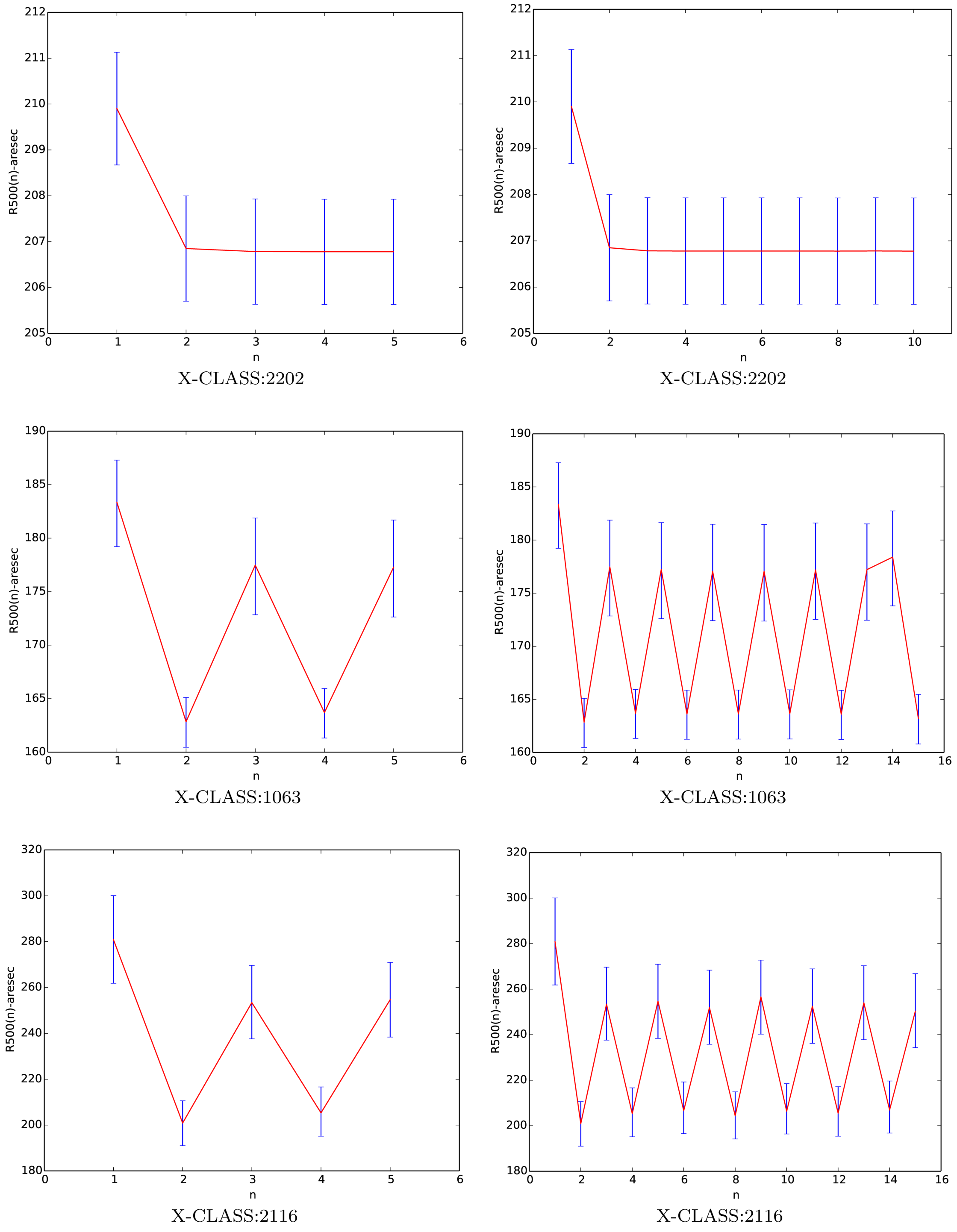

**Figure B1.** Representative graph of $R_{500}$ iteration process. The X-axis represent the frequency of the iteration process, The Y-axis represent the $R_{500}$ in arcsec, and the blue line represent $R_{500}$ result within it's uncertainty. Upper panel(X-CLASS:2202):Left: it shows the results of 5 iteration steps, Right:it shows the results of 10 iteration steps. This panel shows that we can pick any of the $R_{500}$ values since it shows stable behavior and we can ed end the iteration process for that cluster. Middle panel(X-CLASS:1063):Left: it shows the results of 5 iteration steps, Right:it shows the results of 16 iteration steps. This panel shows that we can pick only the value of $R_{500}$ at the point 14, where it did not show stable behavior until after 14 iteration step and then diverge again. Bottom panel(X-CLASS:2116):Left: it shows the results of 5 iteration steps, Right:it shows the results of 15 iteration steps. This panel shows a failure case where we could not be able to find stable value for $R_{500}$